\documentclass[a4paper,fleqn,usenatbib,useAMS]{mnras}

\usepackage{graphicx,multicol}
\usepackage{times}

\usepackage{amsmath} 
\newcommand{\gv}[1]{\ensuremath{\mbox{\boldmath$ #1 $}}} 
 
\renewcommand{\div}[1]{\gv{\nabla} \cdot #1} 
\let\baraccent=\= 
\renewcommand{\=}[1]{\stackrel{#1}{=}} 

\title[PSR J2144$-$3933]{A Single spark model for PSR J2144$-$3933}
\author[Mitra, Basu, Melikidze \& Arjunwadkar]{Dipanjan Mitra$^{1,2}$, Rahul Basu$^{3,2}$, George I. Melikidze$^{2,4}$ \& Mihir Arjunwadkar$^{5}$\\
$^{1}$ National Centre for Radio Astrophysics, Tata Institute of Fundamental Research, Pune 411007, India \\
$^{2}$ Janusz Gil Institute of Astronomy, University of Zielona G\'ora, ul. Szafrana 2, 65-516 Zielona G\'ora, Poland \\
$^{3}$ Inter-University Centre for Astronomy and Astrophysics, Pune, 411007, India; rahulbasu.astro@gmail.com \\
$^{4}$ Abastumani Astrophysical Observatory, Ilia State University, 3-5 Cholokashvili Ave., Tbilisi, 0160, Georgia \\
$^{5}$ Centre for Modeling and Simulation, Savitribai Phule Pune University, Pune 411007, India \\
}

\newcounter{attnctr} 

\usepackage[normalem]{ulem}

\begin{document}



\maketitle

\label{firstpage}
\begin{abstract}

The partially screened vacuum gap model (PSG) for the inner
acceleration region in normal radio pulsars, a variant of the pure
vacuum gap model, attempts to account for the observed thermal X-ray
emission from polar caps and the subpulse drifting timescales.  We
have used this model to explain the presence of death lines, and
extreme location of PSR J2144$-$3933 in the $P-\dot{P}$ diagram. This
model requires maintaining the polar cap near a critical temperature
and the presence of non-dipolar surface magnetic field to form the
inner acceleration region.  In the PSG model, thermostatic regulation
is achieved by sparking discharges which are a feature of all vacuum
gap models.  We demonstrate that non-dipolar surface magnetic field
reduces polar cap area in PSR J2144$-$3933 such that only one
spark can be produced and is sufficient to sustain the critical
temperature.  This pulsar has a single component profile over a wide
frequency range.  Single-pulse polarimetric observations and the
  rotating vector model confirm that the observer's line-of-sight
  traverses the emission beam centrally.  These observations are
consistent with a single spark operating within framework of the PSG
model leading to single-component emission.  Additionally,
single-pulse modulations of this pulsar, including lack of subpulse
drifting, presence of single-period nulls and microstructure, are
compatible with a single spark either in PSG or in general vacuum gap
models.

\end{abstract}

\begin{keywords}
pulsars: general
\end{keywords}

\section{Introduction}
\noindent

The pulsar J2144$-$3933, discovered by \citet{1999Natur.400..848Y},
was the longest period ($P$=8.5 seconds), isolated, radio loud pulsar
for almost two decades until recently when PSR J0250+5854 ($P$=23.5
seconds) was reported by \citet{2018ApJ...866...54T}.  Long period
pulsars are rare: This can be explained by the fact that as pulsars
slow down, they lose their ability to both accelerate and multiply the
pair plasma responsible for radio emission
\citep{1971ApJ...164..529S}.  A rotating neutron star with large
  surface magnetic field ($B$) generates maximum accelerating
potential $\Delta V \propto B/P^2$ external to the star 
\citep[see e.g.][hereafter RS75]{1975ApJ...196...51R}.  The primary pair
plasma, formed due to pair creation in strong surface magnetic field,
is accelerated by a large $\Delta V$.  This leads to further
production of secondary pairs which can form dense clouds of
relativistically streaming pair plasma.  $\Delta V$ reduces
significantly in long-period pulsars and eventually goes below a
critical value where the pair creation process is suppressed.
Coherent radio emission from pulsars is believed to originate in the
growth of instabilities in a dense relativistically streaming pair
plasma (see
e.g. \citealt{1995JApA...16..137M,2002nsps.conf..240U,2002nsps.conf..230L}).
Therefore, the absence of dense pair plasma causes the emission to be
switched off.  It is possible to find a relation between the surface
magnetic field ($B_s$) and the bounding period by considering the
magnetic field configuration and conditions for pair creation in
certain limiting cases.  But observations can find estimates for only 
the dipolar part ($B_d$) of surface magnetic field.  Assuming the
  neutron star to be a homogeneous and uniform sphere of radius $=
  10$ km and mass $= 1.4 M_\odot$ ($M_\odot$ being mass of sun) the 
  dipolar magnetic field can be  estimated as $B_d = 6.2 \times 10^{12} 
  \sqrt{P \dot{P}}$ G, where $\dot{P}$ is the measured pulsar slowdown 
  rate. The surface field magnitude can be related to the dipolar field 
as $B_s = b B_d$, where $b$ is the ratio between the two.  The limiting 
$P$ and $B_s$ relation is expected to be a line in the $P-\dot{P}$ plane 
of the pulsar population; this is usually called the \emph{death line}.  
Hence, any pulsar with physical parameters beyond the death line should 
not be observed as a radio pulsar.

\citet{1999Natur.400..848Y} noted that PSR J2144$-$3933 is an outlier based 
on the then existing death line predictions \citep{1993ApJ...402..264C}.  
The pulsar could only be active at radio frequencies under very specific 
assumptions, either due to extremely complex $B_s$ or a distinctive equation 
of state.  As a result, this pulsar was considered to present significant 
challenges for radio emission theories. This motivated a number of different
approaches including revisiting the pair creation processes in the inner 
magnetosphere \citep{2000ApJ...531L.135Z,  2000ASPC..202..449A,
2001ApJ...550..383G} and exploring the presence of special equation of states 
\citep{2017MNRAS.472.2403Z}.  Pulsar death line estimates are highly 
model dependent and involve several unconstrained parameters, thereby defining 
a broad region in the $P$-$\dot{P}$ diagram instead of a line. This is 
commonly called the \emph{death valley}.  While investigating special 
equations of state, \citet{2017MNRAS.472.2403Z} concluded that a massive 
($>$2$M_\odot$) neutron star is needed to make PSR J2144$-$3933 radio loud. 
Greater mass increases moment of inertia ($I$) which contributes to the 
surface dipolar magnetic field as $B_d \propto \sqrt{I/R^6}$, in turn raising
$\Delta V$ to facilitate pair production.  However, just increasing
$B_d$ is not sufficient to account for the required pair production.
Detailed studies of bright wind nebulae around young pulsars
\citep{deJager2007,Kargaltsev2015} and models for excitation of
coherent radio emission (RS75) require high multiplicity factor
($\kappa \sim 10^5$) between primary and secondary pairs.  $\kappa$
increases with increasing angle between photon direction and $B_s$,
and needs significantly smaller radius of curvature of surface fields
than a purely dipolar field.  Also pair cascade simulations (see e.g.,
\citealt{Timokhin2019}) suggest that such high $\kappa$ can only be
obtained when strongly non-dipolar magnetic fields exist at the
neutron star surface.  The above arguments advocate an increased value
for the ratio $b$ between surface and dipolar fields.

Primarily, there are two classes of models that can produce pair
cascades in the inner magnetosphere of a pulsar.  The first is the
space-charge limited flow (SCLF) where stationary charges can freely
flow from the polar cap.  The second involves non-stationary sparking
discharge from the inner vacuum gap (IVG) where charges cannot escape
the surface due to high binding energy, leading to a high parallel
electric field above the polar cap.  Attempts to account for the
existence of PSR J2144$-$3933 using the SCLF model 
\citep[e.g.,][]{2000ApJ...531L.135Z, 2000ASPC..202..449A} have not 
succeeded in explaining important features of radio emission such as 
subpulse drifting.  On the other hand, sparking discharge from IVG 
models has been more successful in explaining a wide array of radio 
observations. RS75 were the first to suggest that the binding energies 
of surface ions are significantly high for ions to be pulled out from 
the polar cap, thereby facilitating the formation of IVG with strong 
electric fields.  The IVG discharges via a number of localized sparks.  
As these sparks grow both radially and horizontally, a non-stationary
spark-associated pair plasma flow is established along the open
dipolar magnetic field lines.  Radio emission from these
spark-associated flows leads to the observed subpulses.  The growth of
instabilities in the outflowing plasma stream gives rise to coherent
radio emission a few hundred kilometers above the neutron star
surface.  As suggested by RS75 the sparks slowly undergo $\vec{E} 
\times \vec{B}$ drift motions across the IVG which can be observed 
as drifting subpulses.
The electric field in the gap
separates the pair plasma, where the positrons are accelerated upwards
and the backflowing electrons bombard the polar cap surface resulting
in thermal X-ray emission.  \citet[][GM01 hereafter]{2001ApJ...550..383G} 
revisited the formation of IVG in PSR J2144$-$3933 and demonstrated that 
non-dipolar magnetic field with $b \sim 50$ and radius of curvature $R_c 
\sim 10^5$ cm are required.

\citet[][GMG03 hereafter]{2003A&A...407..315G} argued that the
accelerating potential drop in a pure vacuum gap model is very
high, exceeding 10$^{13}$ V.  As a consequence, the predicted drift
speeds of sparks are much faster than the observed ones.
Additionally, the pure vacuum gap model predicts temperatures of about
10$^7$ K from thermal polar cap; these are also much higher compared
to the observational limits.  A high value of the electric field leads to
overestimation of binding energy of ions in the star.  Subsequent
detailed calculations have shown that a strong non-dipolar magnetic
field ($>$10$^{13}$ G) is needed to bind ions to the surface and form
the IVG \citep{1986MNRAS.218..477J,
  2006PhRvA..74f2507M,2006PhRvA..74f2508M}.  This led GMG03 to propose
the partially screened gap (PSG) model which preserves the essential
features of IVG but requires a lower value of electric field due to flow
of low density ions from the surface.  The field anomalies resulting
in strong non-dipolar surface magnetic field, which are essential for
this model, can be generated by Hall-drift instabilities
\citep{2013MNRAS.435.3262G}.  The decreased electric field in the PSG
model naturally accounts for the observed slow drift speeds and lower
polar cap temperatures \citep{2006ApJ...650.1048G}.  The PSG model has
also been used to understand a number of other observational
properties of radio emission, such as the anti-correlation between the
periodicity of subpulse drifting and the spin-down energy loss
\citep{2016ApJ...833...29B}, the phenomenon of mode changing
\citep{2015MNRAS.447.2295S}, etc.

In this paper we intend to understand the emission properties of PSR
J2144$-$3933 within the framework of the PSG model.  In
Sec.\ \ref{sec_obs1} and \ref{sec_obs2} we present highly sensitive
single-pulse observations and radio emission properties of this
pulsar.  In Sec.\ \ref{spark}, we discuss the PSG model in the
presence of a single spark and its implications on the observed radio
emission of PSR J2144$-$3933.

\section{Observations}\label{sec_obs1}

\begin{figure}
\begin{center}
\includegraphics[scale=0.65]{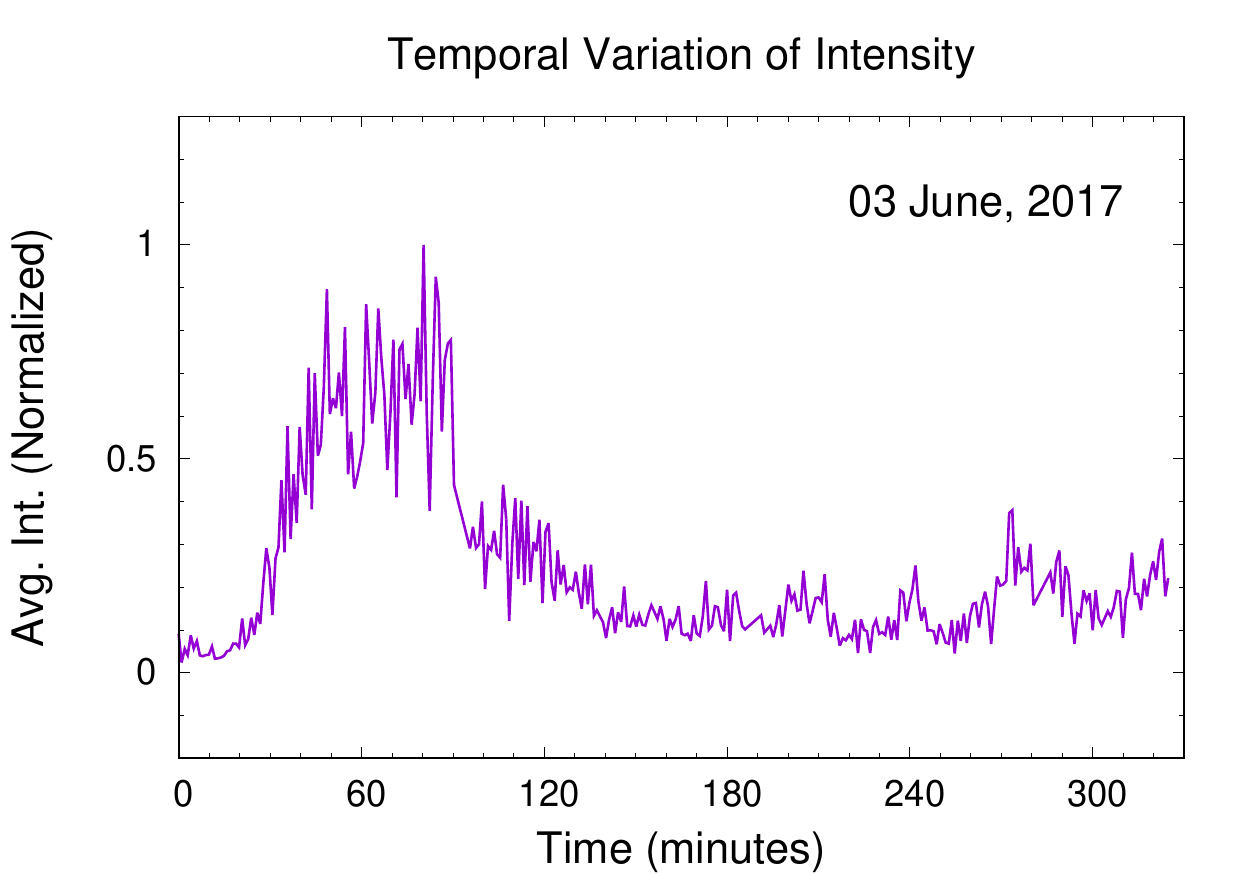}\\
\vspace{20px}
\includegraphics[scale=0.65]{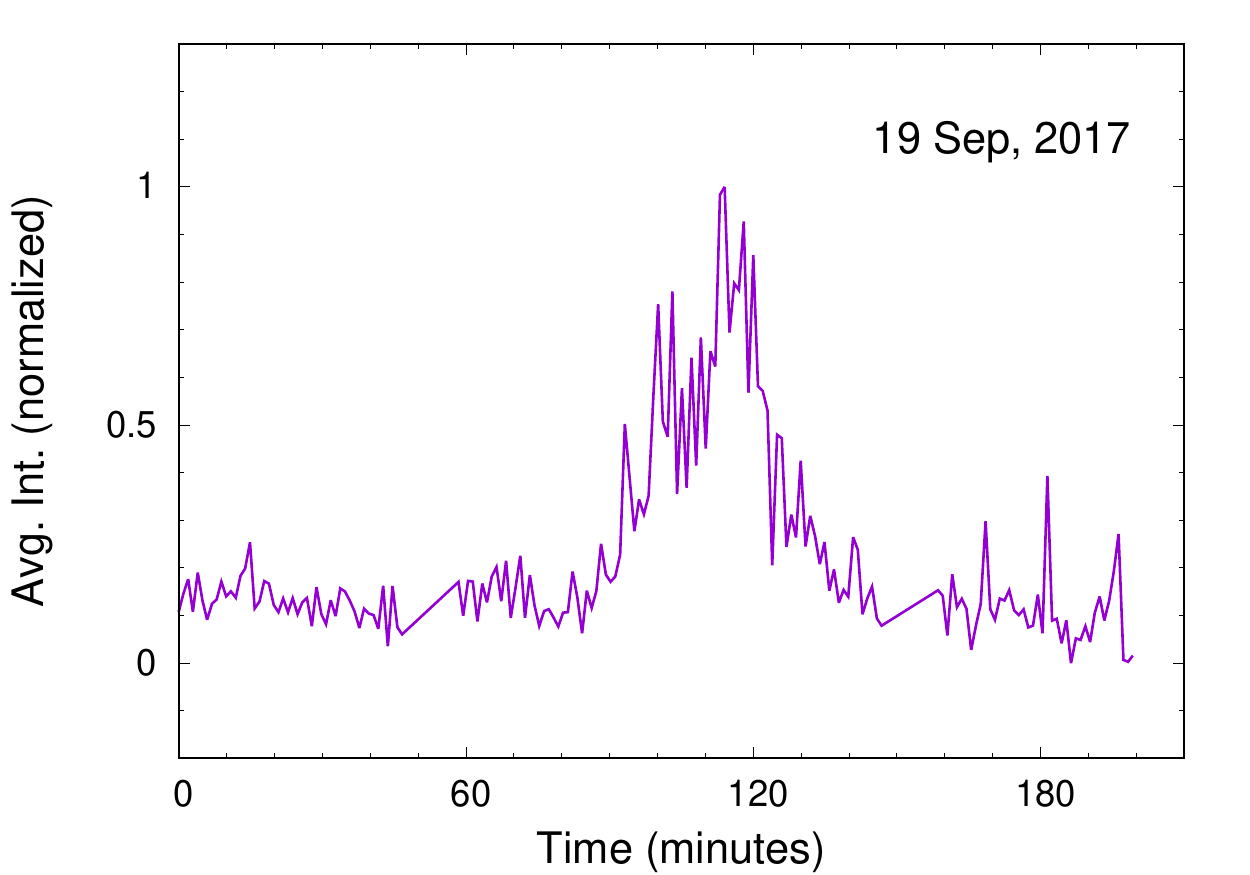}\\
\end{center}
\caption{\label{fig_scint}Variation in the single-pulse intensity of
  PSR J2144$-$3933 over two long observing sessions; top panel: 3 June
  2017, bottom panel: 19 September 2017. Each point in these plots has
  been averaged over about a minute and normalised by the peak
  intensity. Both observations show increased intensities for 50-100
  minute durations. These variations in intensity are most likely a
  result of diffractive scintillations.  }
\end{figure}

\noindent
We have used single-pulse observations from the Giant Meterwave Radio
Telescope \citet[GMRT,][]{1991CuSc...60...95S} to explore the radio
emission properties of PSR J2144$-$3933.  GMRT is an
aperture-synthesis telescope consisting of 30 antennas distributed in
a Y-shaped array with 14 antennas located within a central
square-kilometer area and the remaining 16 spread along its three
arms.  Single-pulse observations using GMRT are usually conducted in
the phased-array mode where signals from the central square antennas
and the first few arm antennas are co-added in phase for increased
detection sensitivities.  The phase offset in each antenna is
determined by observing a nearby high-intensity unresolved source, and
this process is repeated roughly every one or two hours.  The
observations reported here were carried out using the GMRT software
backend \citep[GSB,][]{2010ExA....28...25R} (which has been superseded
by the wideband uGMRT in recent years). Earlier, the GSB supported
observations in several radio frequency bands with 16/32 MHz
bandwidths.  PSR J2144$-$3933 was observed using the GSB as part
of the Meterwavelength Single pulse Polarimetric Emission Survey
\citep[MSPES,][]{2016ApJ...833...28M}.  MSPES conducted high
  sensitivity polarization measurements of single pulses in the
  334 and 618 MHz frequencies using 16 MHz bandwidth.  The time
allotted in this survey allowed recording 325 single pulses at 334 MHz
and 239 single pulses at 618 MHz for this source.  In order to carry
out a detailed study of single-pulse variability, newer observations
of total intensity were conducted on 3 June 2017 and 19 September 2017
at 334 MHz using an increased bandwidth of 33 MHz.  Around 2200 single
pulses were observed on the 3 June 2017 spanning 5 hours on the source
and three phasing intervals in between.  An additional 1500 pulses
were observed on 19 September 2017 lasting for roughly four hours and
two phasing intervals in between.  This newer data was recorded in
filter-bank format with 256 channels and with a time resolution of
122.8 $\mu$sec (finally averaged to $\sim$1.96 msec).  Post-processing
of the data involved removal of spectral channels affected by
radio-frequency interference, correcting the dispersion spread across
channels before averaging across the entire band, and smoothing over
variations across the baseline, as detailed in
\citet{2016ApJ...833...29B}.

The observed single pulses showed large intensity variations in both
observing sessions (Fig.\ \ref{fig_scint}), with high-intensity
emission was seen for roughly 50-100 minutes followed by
lower-intensity signals.  PSR J2144$-$3933 is a relatively nearby
source with an estimated dispersion measure DM $= 3.35$ pc cm$^{-3}$
and distance $d = 0.16$ Kpc 
\footnote{Parameters used from ATNF Pulsar Catalogue:
  \url{https://www.atnf.csiro.au/research/pulsar/psrcat/},
  \citet{2005AJ....129.1993M}}
  \citep{1996MNRAS.279.1235M,1998MNRAS.297...28D}.  The observed
intensity variations are most likely a result of diffractive
scintillations which are propagation effects due to interaction of the
pulsar emission with the interstellar medium
\citep{1990ARA&A..28..561R}.  The primary variabilities associated
with diffractive scintillations are the decorrelation bandwidth
($\Delta\nu_{\mbox{\textsc{diss}}}$) and the diffractive timescale
($\tau_{\mbox{\textsc{diss}}}$) which can be estimated approximately
using pulsar parameters.  Using $\Delta\nu_{\mbox{\textsc{diss}}}$ =
11 MHz ($\nu$/GHz)$^{4.4}$(d/Kpc)$^{-2.2}$
\citep{1985ApJ...288..221C}, the estimated bandwidth is $\sim$4 MHz at
$\nu =0.334$ GHz and should be averaged over the observing bandwidths.
The diffractive timescale is given as $\tau_{\mbox{\textsc{diss}}}$ =
(A/$V_{\mbox{\textsc{iss}}}$)
($\Delta\nu_{\mbox{\textsc{diss}}}$/MHz)$^{0.5}$($\nu$/GHz)$^{-1}$ s
\citep{1998ApJ...507..846C}, where $V_{\mbox{\textsc{iss}}}$ is the
relative velocity of the pulsar.  The transverse velocity of PSR
J2144$-$3933 is around 120 km~s$^{-1}$ which makes the diffractive
timescale to be $\sim$30 mins.  The estimated timescale is consistent
with the intensity variations seen in Fig.\ \ref{fig_scint}.

\section{Radio Emission Features} \label{sec_obs2}

\subsection{Stokes-I Single Pulse Modulation: Nulling and Subpulse Drifting} \label{sec_Imod}

\begin{figure*}
\begin{center}
\begin{tabular}{cc}
\mbox{\includegraphics[scale=0.36]{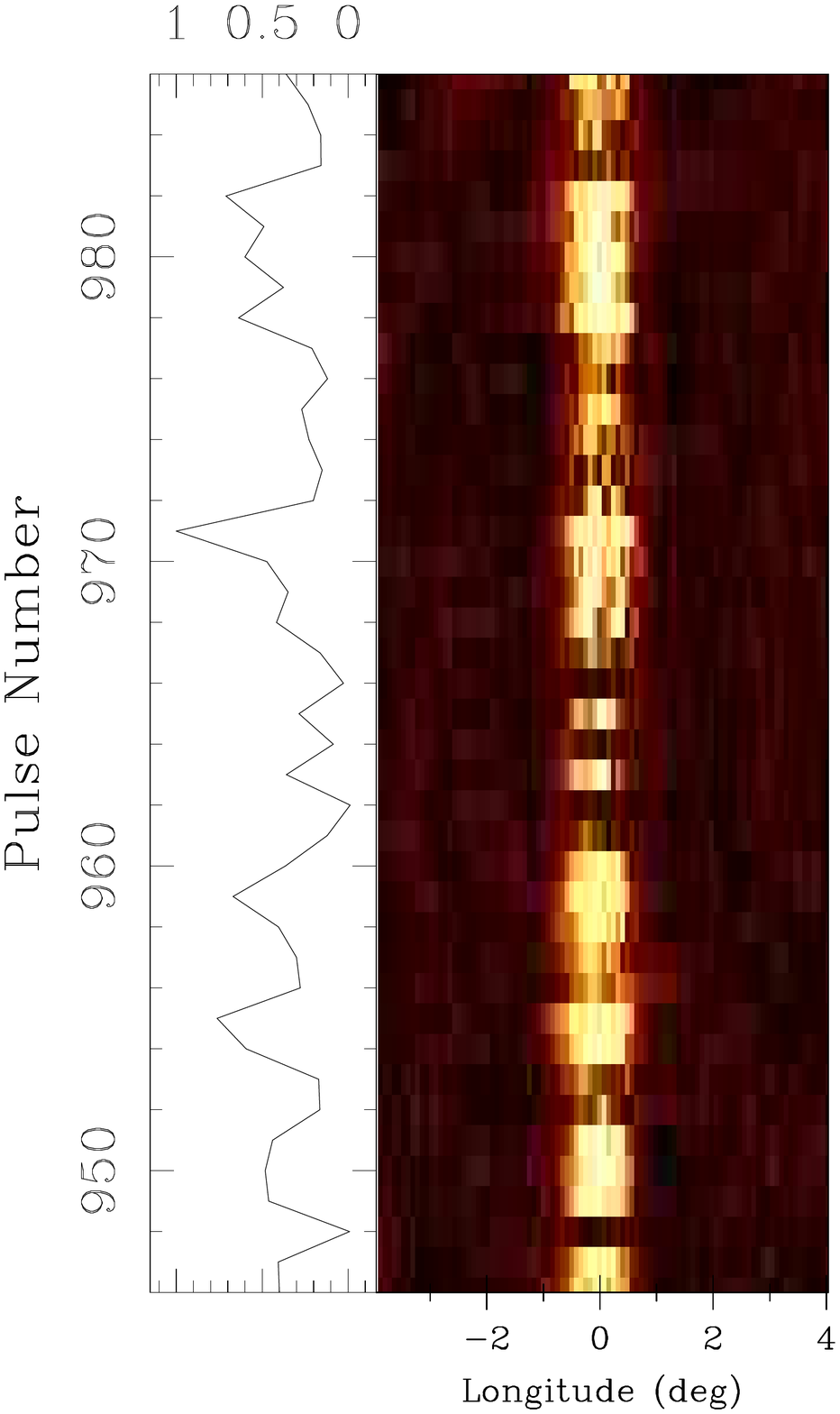}} &
\hspace{60px}
\mbox{\includegraphics[scale=0.58]{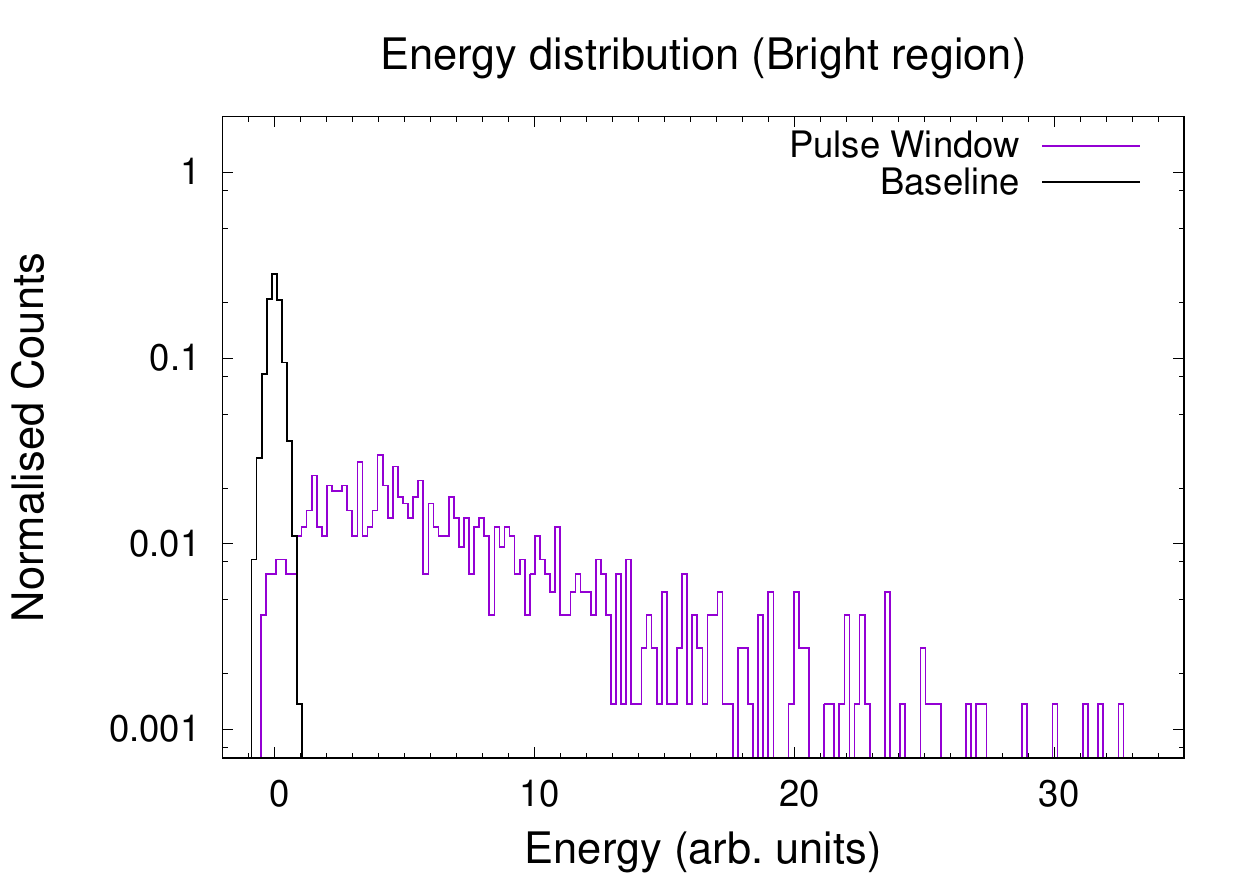}} \\
\end{tabular}
\caption{\label{fig_null}Nulling behaviour in PSR J2144$-$3933. Left
  panel: a short section of $\sim$40 pulses between pulse \#945 and
  \#985 from the start of the observation (3 June 2017) where
  single-period null pulses can be clearly seen. Right panel: Energy
  distribution (violet) of the pulse window corresponding to the
  bright interval between pulse \#350 and \#1100 during the same
  observing session, together with the noise distribution (black) of
  the baseline in the off-pulse region. The null pulses show a
  separate distribution overlapping the baseline histogram.}
\label{fig_null}
\end{center}
\end{figure*}

The two primary single-pulse modulations intrinsic to pulsar radio
emission are the phenomena of nulling and subpulse drifting.  Nulling
is the sudden disappearance of emission, and ranges from short
durations (few periods) to much longer stretches (hours at a time).
Subpulse drifting, as discussed earlier, is the periodic modulation of
subpulses within the pulse window.  Nulling in PSR J2144$-$3933 was
observed in MSPES, but no detailed analysis was carried out due to the
relatively few pulses in the survey.  Fig.\ \ref{fig_null} (left
panel) shows a short pulse sequence from the observations on 3 June
2017 which shows single-period nulls.  In order to characterise the
nulling behaviour of this pulsar, we estimated the pulse-energy
distributions from average energy across the pulse window along with
the baseline distribution from the off-pulse region
\citep{1976MNRAS.176..249R}.  Nulling results in a bimodal
distribution of pulse energies where null pulses are overlapping with
the baseline histogram.  The nulling fraction identifies the relative
abundance of nulling and is estimated as the ratio between the peaks
of the null and the baseline distributions.  As discussed earlier, the
single-pulse intensities show large fluctuation in PSR J2144$-$3933
due to interstellar scintillations.  In order to characterise the
nulling behaviour we considered only the duration when the pulsar was
in the bright state.  This corresponded to around 700 pulses in the
pulse range \#350-\#1100 for the 3 June 2017 observations whose energy
histograms are shown in Fig.\ \ref{fig_null} (right panel).  The null
pulses are clearly seen separated from the burst distribution and we
estimated the nulling fraction to be around 3.5\%.  In recent years,
the presence of periodicity associated with nulling has been reported
in a number of pulsars \citep{2009MNRAS.393.1391H}.  A standard method
to estimate periodic nulling behaviour has been developed by
\citet{2017ApJ...846..109B}, where the null and bright pulses are
replaced by zero and unity respectively, and then this binary series
is Fourier-transformed.  A similar study was conducted for the pulse
sequence during the bright phase of PSR J2144$-$3933 as shown in
Fig.\ \ref{fig_perd} (left panel).  No clear nulling-related
periodicity is seen for this pulsar.

\begin{figure*}
\begin{center}
\begin{tabular}{cc}
\mbox{\includegraphics[scale=0.36]{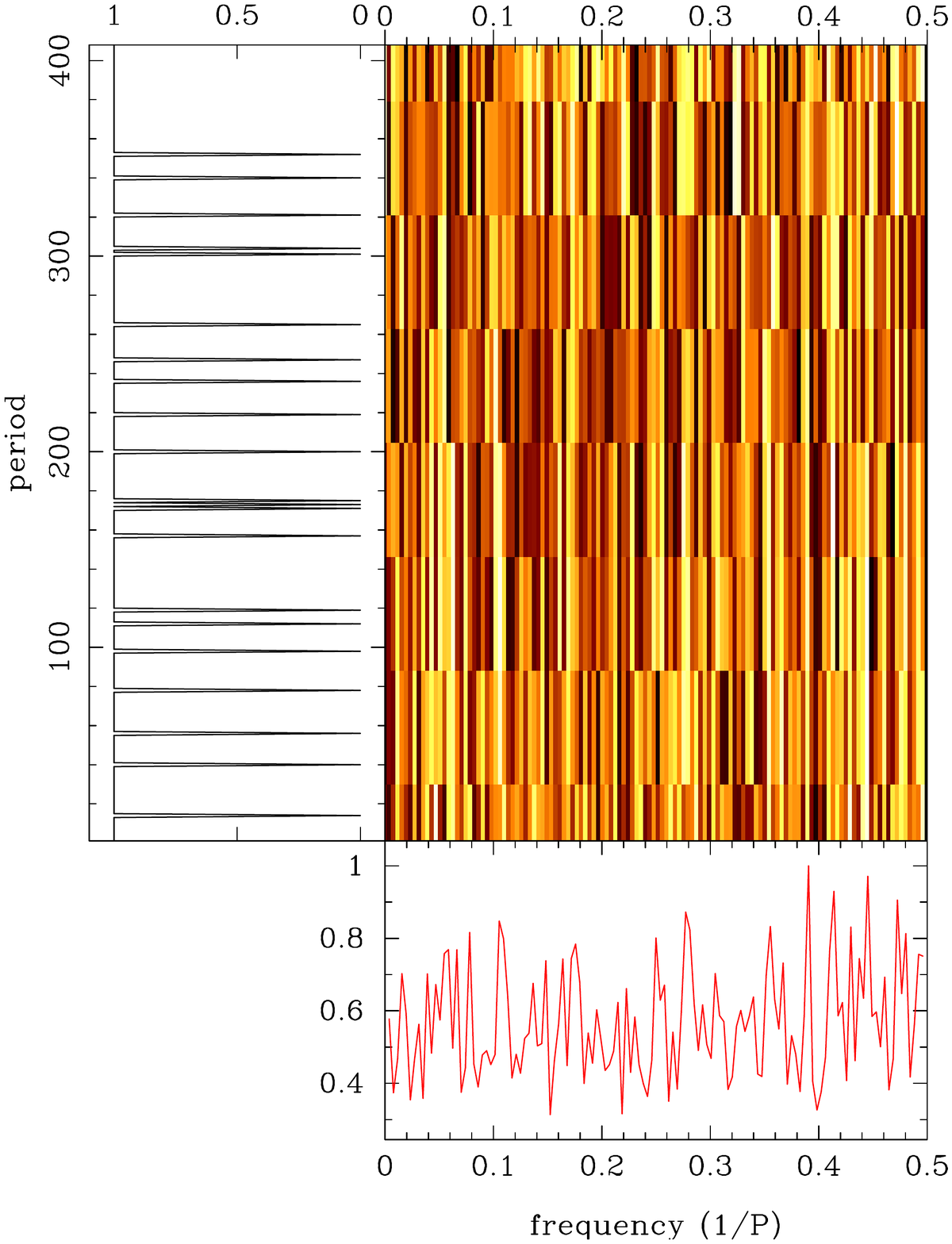}} &
\hspace{40px}
\mbox{\includegraphics[scale=0.36]{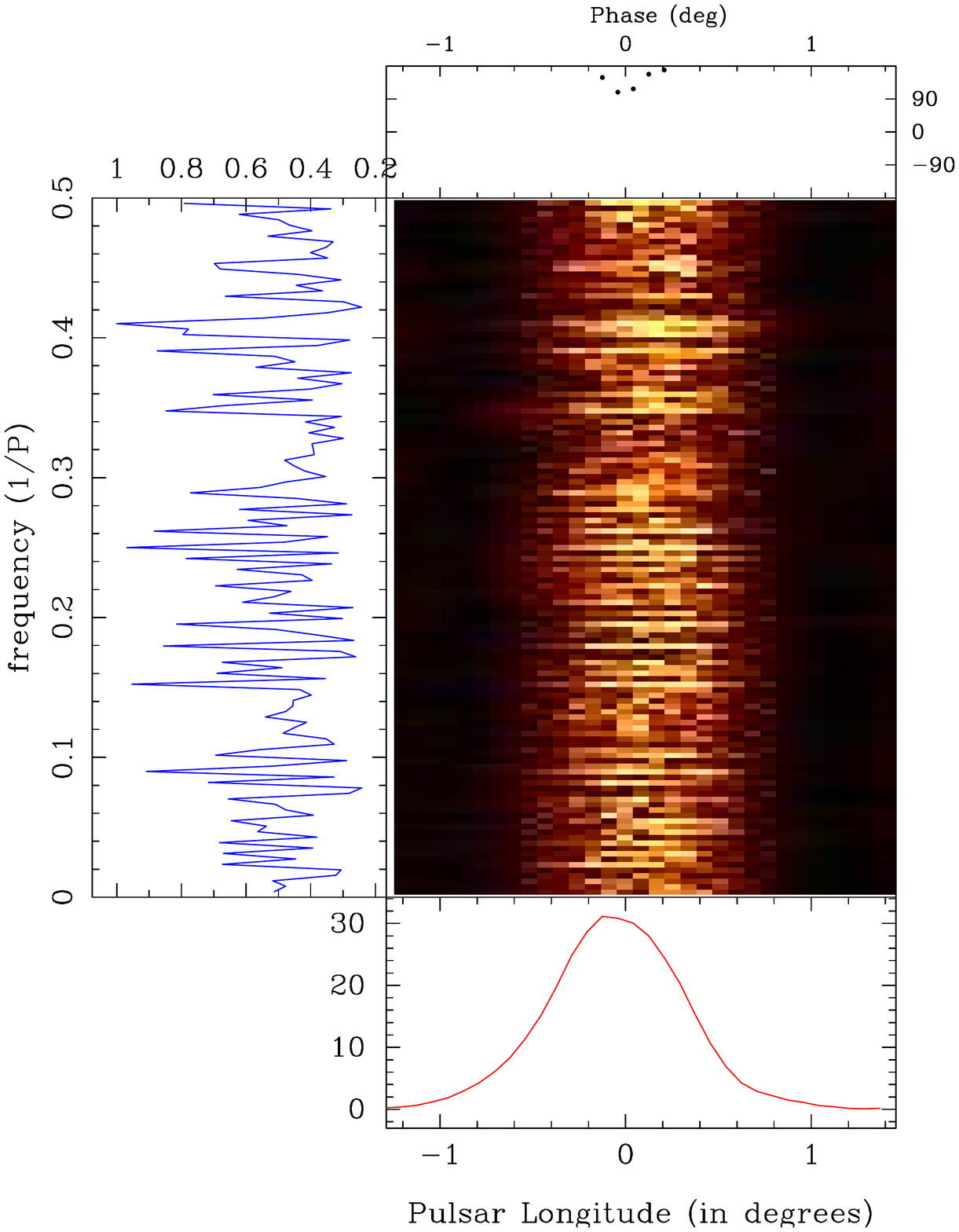}} \\
\end{tabular}
\caption{\label{fig_perd}Periodic behaviour in single pulse sequences
  of PSR J2144$-$3933. Left panel: FFT power spectrum of the binary
  null sequence showing no signs of dominant periodicity/ies. Right
  panel: Longitude-resolved fluctuation spectrum (LRFS) for this
  pulsar showing no dominant peak (except at 0 frequency), indicating
  the absence of subpulse drifting.}
\label{fig_perd}
\end{center}
\end{figure*}

We have also investigated the presence of subpulse drifting by using
the technique of longitude-resolved fluctuation spectra
\citep[LRFS,][]{1973ApJ...182..245B}.  This involves carrying out
Fourier transforms along each longitude bin in the pulse window.  Any
periodicity will be seen as a peak in the Fourier spectrum.  We have
estimated the LRFS during the bright emission state of PSR
J2144$-$3933 for the two long observing sessions.  The LRFS was
estimated for a number of different intervals ranging from 250 to 600P
at a time on both observing sessions.  An example of LRFS between
pulse number 460 and 716 from the start of the observing session on 3
June is shown in the right panel of Fig.~\ref{fig_perd}.  We did not
detect any periodic behaviour in this pulsar: This suggests the
absence of subpulse drifting.  It should be noted that any periodic
behaviour longer than $\tau_{\mbox{\textsc{diss}}}$ (i.e., around
300-400$P$) will be hidden by scintillations and hence not detectable.

\subsection{Polarization and Geometry}\label{sec_geom}

\begin{figure}
\begin{center}
\includegraphics[scale=0.35,angle=0.]{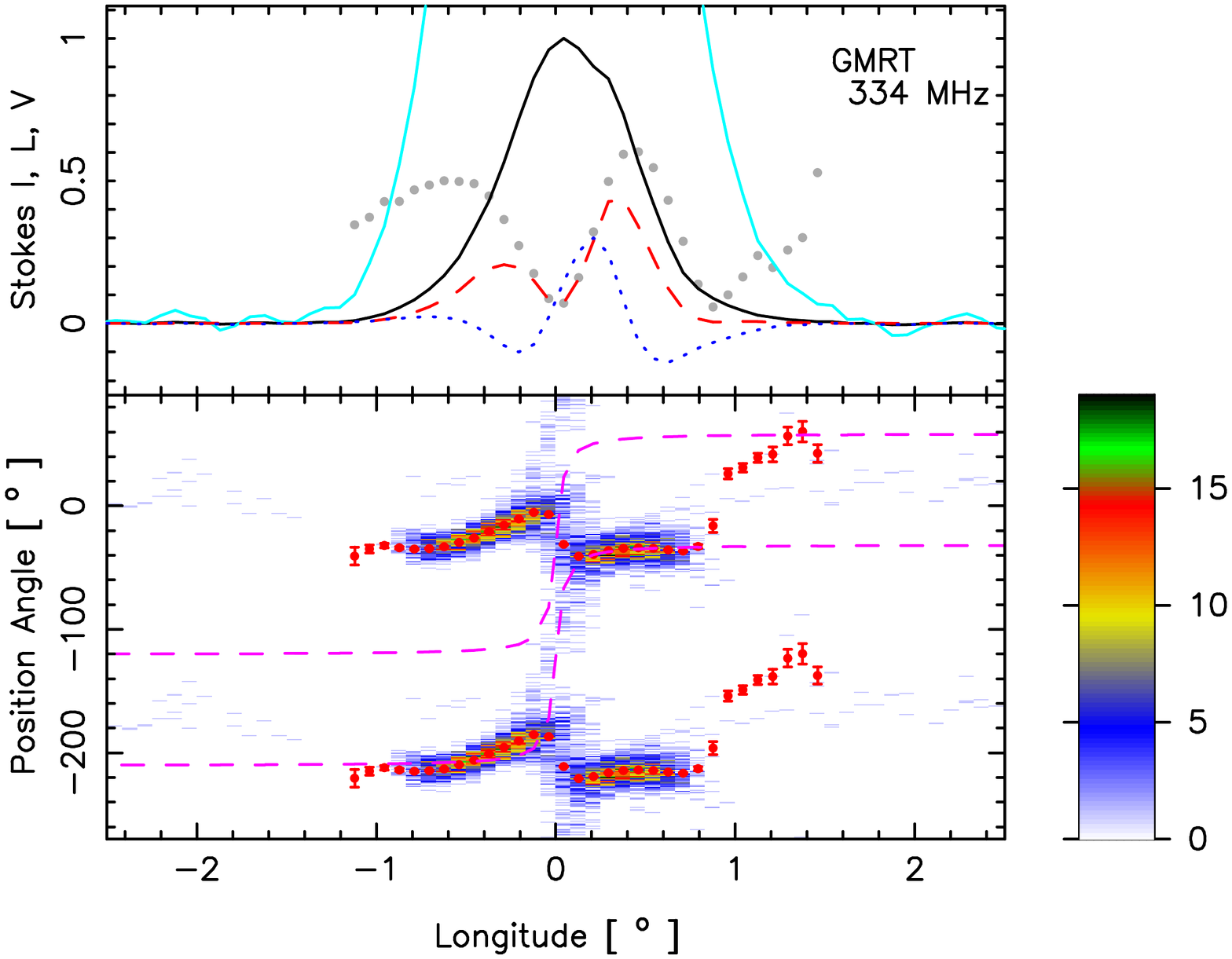}
\caption{\label{fig_pol}Polarization properties of PSR J2144$-$3933 at
  334 MHz. Upper panel: Total intensity (Stokes $I$; solid curve),
  total linear polarization ($L$=$\sqrt{Q^2+U^2}$; dashed red),
  circular polarization (Stokes $V$; dotted blue), the fractional
  linear polarization ($L/I$; gray points), and a zoomed version of
  the total intensity signal ($10\times I$; cyan curve). Lower panel:
  The colour scale shows the single polarization position angle (PPA)
  contour. The average PPA traverses are plotted in red.  The rotating
  vector model (RVM) has been used to fit the PPA traverse and is
  plotted twice for the two polarization modes (magenta dashed) along
  with the PPA. The basic parameters corresponding to the RVM fit are
  $\alpha = 72^{\circ}$, $\beta = 0.03^{\circ}$, $\phi_{\circ} =
  0.0^{\circ}$ and $\Psi_{\circ} =-119^{\circ}$, with the origin at
  the fitted PPA inflection point (see Sec.\ \ref{sec_geom} for
  details).}
\end{center}
\end{figure}

\noindent
Average profiles of PSR J2144$-$3933 are available at frequencies
ranging from 334 MHz to 1.4 GHz and show a single-gaussian like shape
whose width remain roughly constant for all frequencies.  The
estimated width measured between the outer points with 50\% peak
intensity ($W_{50}$) is around 0.8$^{\circ}$.  The polarization
observations have been reported at 334 and 618 MHz in MSPES and at 667
MHz by \citet{1998MNRAS.295..280M}.  In all cases, the pulsar is seen
to be strongly polarized, with percentage linear and circular
polarizations to be around 30\% and 5\% respectively.
Fig.\ \ref{fig_pol} presents the average and single-pulse polarization 
position angle (PPA) at 334 MHz from MSPES. For each single pulse the PPAs are 
estimated at longitudes where the linear polarization power is more than twice
the noise rms in the off pulse region. The PPA distribution from all relevant 
pulses is averaged within a $1\degr \times 1\degr$ cell and shown as a colour
plot in bottom panel of Fig.\ref{fig_pol} (with relevant colour scale shown 
at the lower right of the figure). The average PPA is overplotted in red over 
this colour contour. Note that the PPA exhibits a fast
swing across the profile.  The circular polarization changes sign near
the profile center as well as as towards the trailing edge of the
profile.  Based on these properties \citet{1998MNRAS.295..280M}
suggested that the line of sight in this pulsar cuts the central part
of the emission beam.

Alternatively, the availability of highly sensitive polarization
measurements from MSPES makes it possible to assess the line of sight
geometry by fitting the rotating vector model (RVM) to the PPA
traverse.  According to the RVM, as the star rotates and the line of
sight traverses the emission region, the angle ($\Psi$) between the
projected line of sight and the magnetic field vector changes as a
function of pulse phase.  For a star-centered dipolar magnetic field
with angle $\alpha$ between the rotation axis and the magnetic axis,
and angle $\beta$ between the rotation axis and the observer's line of
sight, the RVM has a characteristic S-shaped traverse given by
\begin{equation}
\Psi = \Psi_{\circ} + \\\tan^{-1} \left( \frac{\sin(\alpha)\sin(\phi-\phi_{\circ})}
{\sin(\alpha + \beta)\cos(\alpha) - \sin(\alpha)\cos(\alpha+\beta)
\cos(\phi-\phi_{\circ})}\right)
\label{req1}
\end{equation}
where $\Psi_{\circ}$ and $\phi_{\circ}$ are the arbitrary phase
offsets for the position angle $\Psi$ and longitude $\phi$
respectively.  We have examined the possibility of fitting the PPA
traverse of PSR J2144$-$3933 using the RVM.  A jump of around
50$^{\circ}$ is seen in the PPA around zero longitude which is also
associated with a substantial dip in the linear polarization
(Fig.\ \ref{fig_pol}).  In the longitude range between 0$^{\circ}$ and
1$^{\circ}$, the PPA is continuous but shows a jump of around
80$^{\circ}$ beyond the 1$^{\circ}$ longitude.  We fit the RVM by
considering the PPA traverse in the longitude range between $-1^{\circ}$
to $1^{\circ}$ to correspond to one polarization mode, with
phase-wrapping around 0$^{\circ}$, while the extreme jump in PPA near
the trailing edge corresponds to the orthogonal polarization mode.  We
have obtained a reasonable fit to the RVM but, as noted in earlier
studies (e.g.,
\citealt{1997A&A...324..981V,2001ApJ...553..341E,2004A&A...421..215M}),
the $\alpha$ and $\beta$ values obtained from these fits are highly
correlated and unreliable.  In Fig.\ \ref{fig_pol}, the magenta curve
in the lower panel shows the RVM fits for the geometry specified by
$\alpha = 72^{\circ}\pm 10^{\circ}$ and $\beta =
0.03^{\circ}\pm0.01^{\circ}$.  The longitude phase offsets have been
subtracted and have errors of $\phi_{\circ} = 0.0^{\circ} \pm
0.5^{\circ}$ and $\Psi_{\circ} = -119^{\circ} \pm 5^{\circ}$.  The
choice of this specific geometry can be justified as follows.

The inflexion point or steepest gradient (SG) point of the RVM occurs
at the phase $\phi_{\circ}$, with the SG point being significantly
better constrained by the RVM fits.  The relation between different
geometrical angles at the SG point is obtained from Eq.\ \ref{req1} as
\begin{equation}
\sin(\alpha)/\sin(\beta)
= \mid d\Psi/d\phi \mid_{\mbox{max}}
\label{req2}
\end{equation}
In PSR J2144$-$3933 the slope has a large value $\mid d\Psi/d\phi \mid
\sim 1.8\times 10^3$.  This suggests that $\beta$ is extremely close
to zero, and hence the observer cuts the pulsar beam centrally.  The
shape of the pulsar radio beam is most commonly represented by a
central core emission surrounded by nested conal structure.  The
observed profile shapes with different components and classifications
depend on the pulsar period, period derivative, and the line-of-sight
geometry (see \citealt{
  1983ApJ...274..333R,1993ApJ...405..285R,1999A&A...346..906M,
  2000ApJ...541..351G}).  Several careful studies have shown that the
distribution of component width $W_{50}$ with respect to period has a
lower boundary line which scales as $P^{-0.5}$
\citep{1990ApJ...352..247R,2018ApJ...854..162S}.  This scaling is
similar to the opening angle of dipolar magnetic field lines.  In
pulsars like J2144$-$3933, with central cuts of the emission beam, the
component associated with the SG point is identified as the core
emission.  It has been argued by \citet{1990ApJ...352..247R} that
$W_{50}$ of the core at 1 GHz is related to $\alpha$ as $\sin\alpha =
2.45 P^{-0.5}/W_{50}$.  \citet{2018ApJ...854..162S} confirmed the
$P^{-0.5}$ scaling at lower frequencies and found the relation to be
$\sin\alpha = 2.39 P^{-0.5}/W_{50}^{0.3GHz}$ at 0.3 GHz.  The
$W_{50}^{0.3GHz}$ of PSR J2144$-$3933 is 0.86$^{\circ}$ which gives
$\alpha \sim 72^{\circ}$.  This justifies our choice of $\alpha$ in
the RVM fits.
\subsection{Quasiperiodic structure in Single Pulses}

\begin{figure}
\begin{center}
\mbox{\includegraphics[scale=0.3,angle=0.]{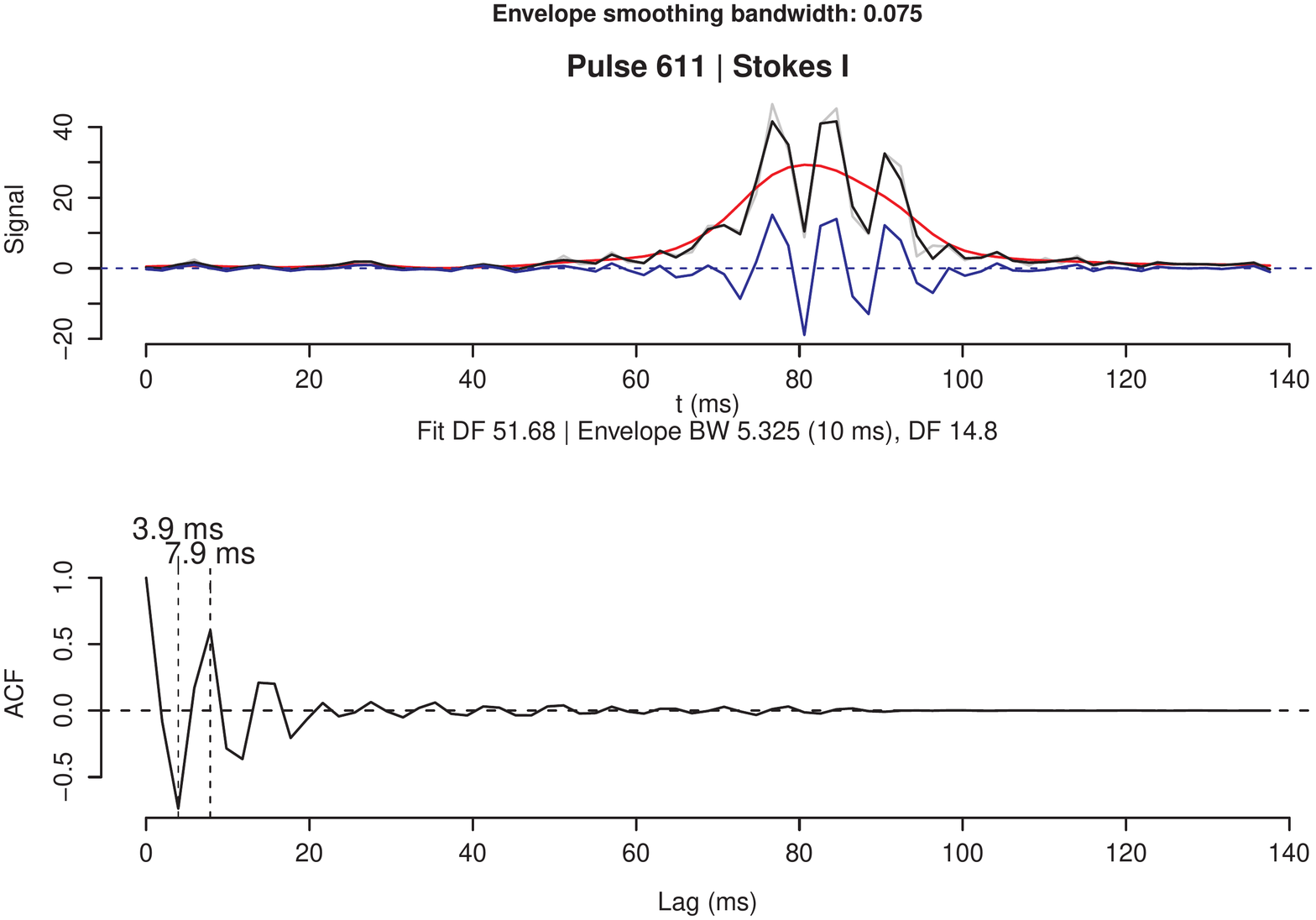}} \\
\vspace{30px}
\mbox{\includegraphics[scale=0.3,angle=0.]{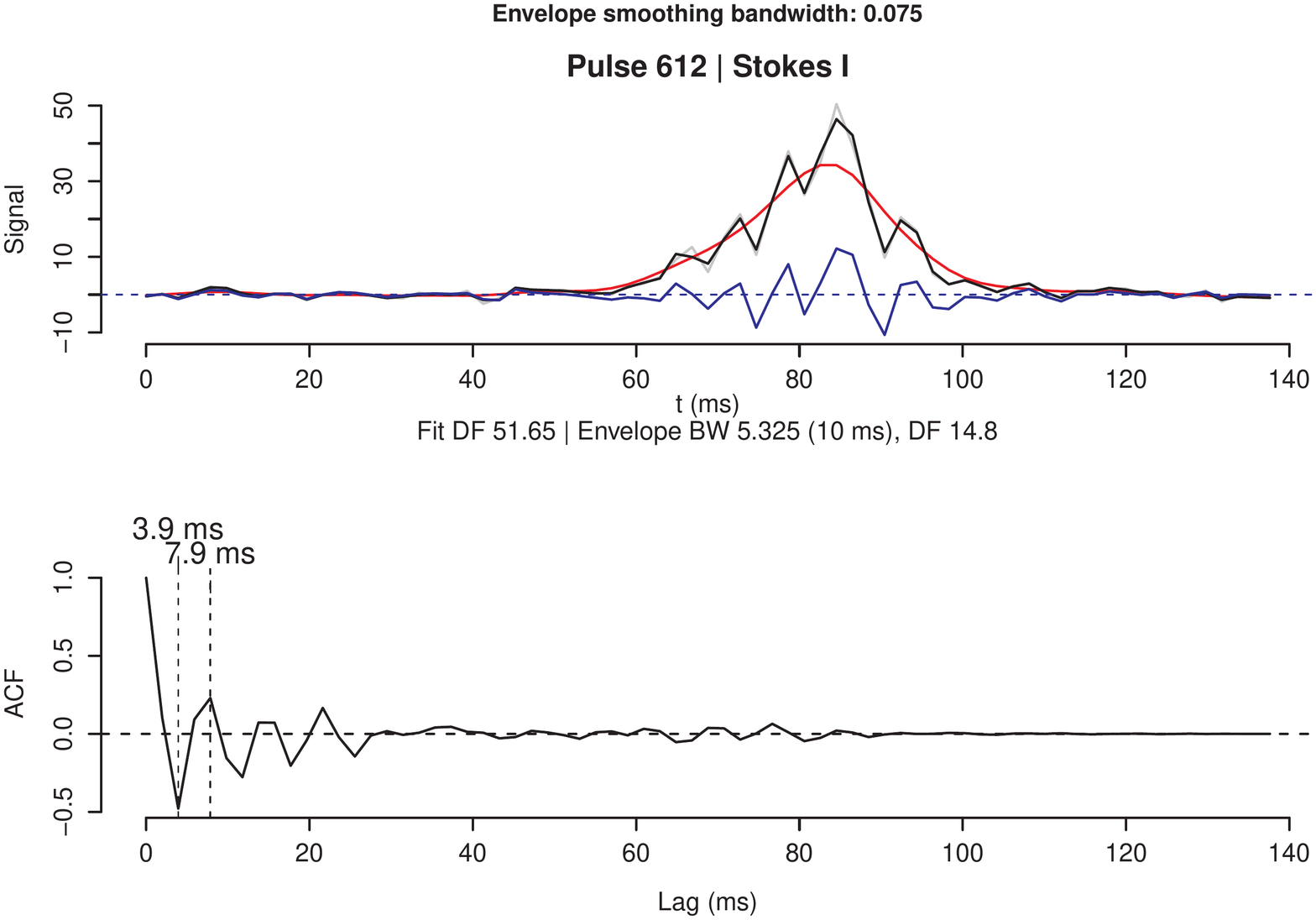}} \\
\vspace{30px}
\mbox{\includegraphics[scale=0.3,angle=0.]{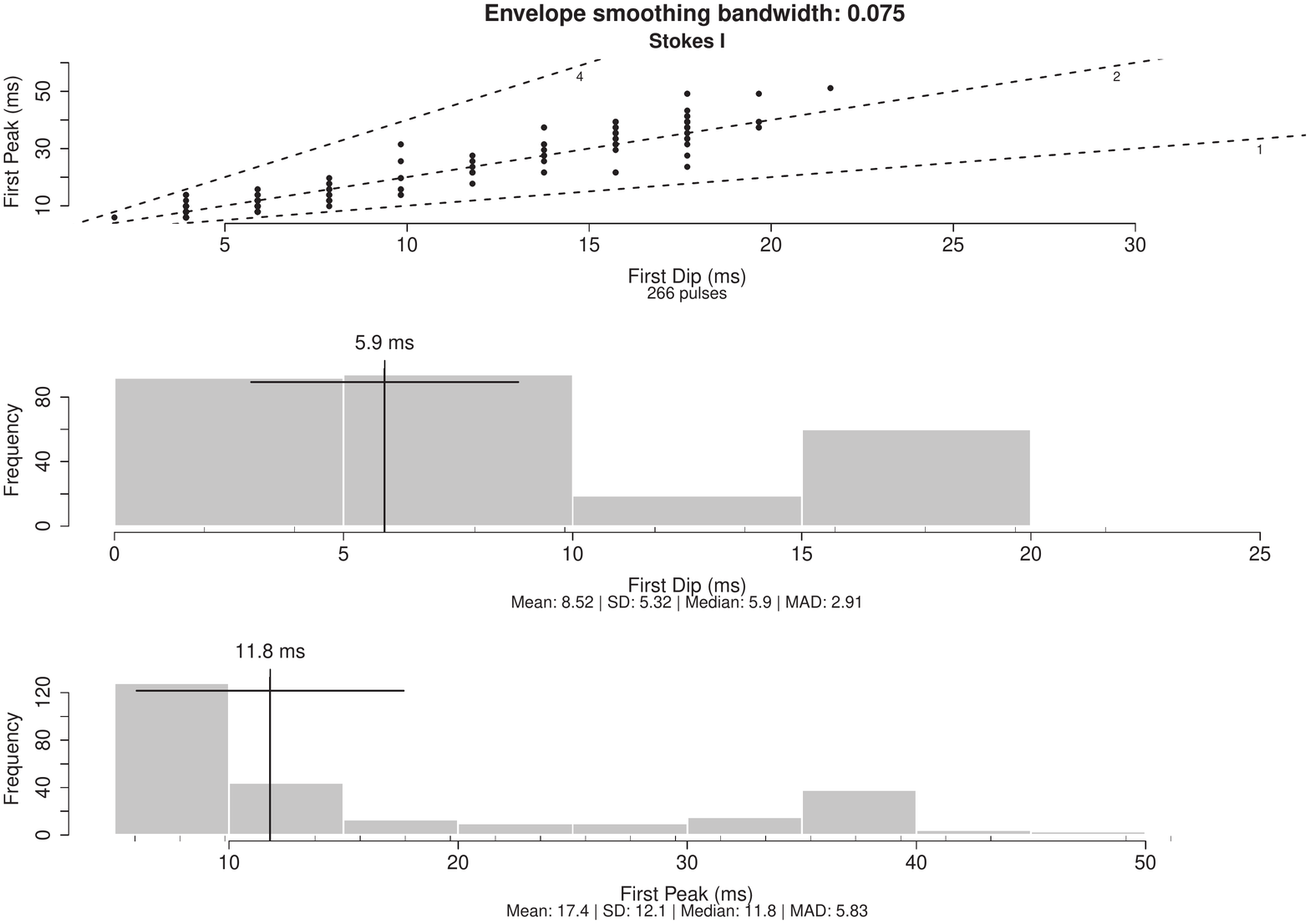}} \\
\caption{\label{fig_micro}Analysis scheme used to infer the timescale
  of quasiperiodic structures in the total intensity single pulses of
  PSR J2144$-$3933 at 334 MHz observing frequency. The top and middle
  plots show two examples of these features, where the technique of
  \citet{2015ApJ...806..236M} has been applied using a smoothing
  bandwidth $0.075$. The lower panel in the bottom plot represents the
  histograms of the estimated time scales $P_{\mu}$ of the
  microstructures, similar to Fig.\ 6 in
  \citet{2015ApJ...806..236M}. The upper panel shows the relation
  between the first dip and the first peak in the autocorrelation
  function (ACF), while the middle panel shows the distribution of the
  first dip in the ACF.}
\end{center}
\end{figure}

\noindent
The subpulses in certain cases exhibit quasiperiodic structures
superposed on top of a smooth emission envelope.  These structures
have a characteristic periodicities ($P_{\mu}$) ranging from hundreds
of $\mu$sec to several msec.  This quasiperiodic behaviour, commonly
known as microstructure, was first observed by
\citet{1968Natur.218.1122C} and has subsequently been seen in a large
number of pulsars.  A detailed and careful analysis of this phenomenon
was carried out by \citet{2015ApJ...806..236M} using a new method.
This study includes pulsars with a wide range of periods between 100
msec and 3.5 sec, and detected such quasiperiodic structures in all
four stokes parameters.  The study also found a tendency of $P_{\mu}$
to increase with the pulsar period $P$ approximately as $P_{\mu} (ms)
\sim 1.3\times10^{-3} P$.

The presence of quasiperiodic structures is clearly visible in
high-sensitivity total intensity measurements of single pulses of PSR
J2144$-$3933 as shown in the top and middle plots of
Fig.\ \ref{fig_micro}.  We have carried out a detailed analysis of the
microstructures using the analysis scheme developed by
\citet{2015ApJ...806..236M}.  The estimated $P_{\mu}$ is 7.9 msec
using smoothing bandwidth $0.075$ for the two single pulses shown in
Fig.~\ref{fig_micro}.  In order to estimate the distribution of
$P_{\mu}$ we identified 355 single pulses with signal to noise ratio
greater than 15 for which microstructure analysis could be carried
out.  A detailed quantitative analysis leads to the median $P_{\mu}=
11.8\pm6$ msec using the smoothing bandwidth $0.075$
\citep[Fig.~\ref{fig_micro} bottom plot; see
  also][]{2015ApJ...806..236M}.  Our estimate of the microstructure
timescale ($P_{\mu}= 11.8\pm6$ msec) for this pulsar ($P=8.5$ sec) is
in good agreement with the consensus in the field ($P_{\mu} = 11$
msec) \citep[see also][]{1979AuJPh..32....9C, 2002MNRAS.334..523K}.

\section{The Single spark model of PSR J2144$-$3933}\label{spark}
\noindent
The polarization behaviour of PSR J2144$-$3933 suggests that its radio
emission originates from open-line region of the dipolar magnetic
field, with the observer's line of sight cutting the radio emission
beam centrally, and emission comprising of a single component akin to
the core emission.  The presence of single-period nulls and the
microstructure periodicity are both consistent with the emission
behaviour of normal pulsars.  A number of detailed observations over
the years have provided strong evidence that coherent radio emission
in normal pulsars is excited by curvature radiation from charge
bunches, which detaches from the pulsar magnetosphere at heights below
10\% of the light-cylinder radius \citep[for a review,
  see][]{2017JApA...38...52M}.  Recent theoretical developments have
also demonstrated that nonlinear growth of Langmuir waves in
relativistically-moving pair plasma can result in the formation of
such stable charge bunches which are capable of exciting curvature
radiation in pulsars
\citep{2000ApJ...544.1081M,2004ApJ...600..872G,2009ApJ...696L.141M,
  2014ApJ...794..105M,2018MNRAS.480.4526L}.  This essentially requires
the presence of an inner acceleration region where a non-stationary
relativistic flow of pair plasma can be established.  A prototype
for inner acceleration region is the IVG model of RS75.
The IVG discharges in the form of several isolated sparks
which are responsible for generating non-stationary flow of
relativistic pair plasma.  The growth of instabilities in the
spark-associated plasma columns excite coherent curvature radiation at
radio emission regions and are observed as subpulses in pulsars.  The
single-component emission in PSR J2144$-$3933 arises due to a single
plasma column, where the polar cap size is such that only one spark
can be accommodated.  If the size of the actual polar cap is smaller
than the fully-developed spark then the pulsar will not be radio loud.
This idea was previously explored by GM01 \citep[see 
also][]{2000ApJ...531L.135Z} within the IVG model (RS75).  We will first 
review the ideas of GM01 before exploring the implications of single-spark 
in the PSG model.

There are two primary requirements for the generation of radio
emission in RS75 model.  The first is the formation of an IVG, and
the second is facilitating conditions in IVG to generate
copious amounts of pairs, their subsequent acceleration, and the
cascading of pairs.  The formation of the IVG is related to the
problem of binding energies of electrons and ions on the stellar
surface.  Depending on whether $\vec{\Omega}\cdot\vec{B}> 0$ or
$\vec{\Omega}\cdot\vec{B}< 0$, the polar cap should either be
negatively charged or positively charged.  RS75 argued that only in
the case of $\vec{\Omega}\cdot\vec{B}<$ 0 the positive ions cannot be
extracted from the stellar surface (because their binding energies are
much greater than that of electrons) and the IVG can be formed.  The
binding energy of ions depends on the temperature of the polar cap
surface as well as the surface magnetic fields.  In their seminal
work, RS75 overestimated the value of the binding energy and hence
their estimates of potential drop above the polar cap was extremely
high.  Subsequently, improved calculations of binding energy were
available which included the dependence on the strength of the surface
magnetic fields
\citep{1986MNRAS.218..477J,2006PhRvA..74f2507M,2006PhRvA..74f2508M}.
In the literature, the estimates from \citet{1986MNRAS.218..477J} are
widely used.  Here, an empirical relationship between the equivalent
temperature ($T_i$) of the binding energy of ions and the surface
magnetic field ($B_s$) is given to be $T_i =
10^6(B_s/B_{q})^{0.73}\rm{G}\approx
(1.2\times10^5)b^{0.7}(P\dot{P}_{15})^{0.36}$ G,where $B_q
=m^2c^3/e\hbar = 4.4\times10^{13}$ G is the critical magnetic field at
which the electron gyro-frequency equals its rest mass.  Thus, for
strong $B_s$, there exists a critical temperature ($T_i$) such that
ions cannot escape from surface and the IVG can be formed if the polar
cap surface temperature $T_s \leq T_i$.  The second condition for
radio emission is related to the production of copious pairs and
formation of sparks in the IVG.  RS75 showed that the accelerating
potential $\Delta V$ in the IVG is given by $\Delta V = 2 \pi (B_s/P)
h^2$ where $h$ is the height of the gap.  If $\Delta V$ falls below a
critical value, sparks cannot be produced and, therefore, pulsar
emission will be switched off.  RS75 argued that $h$ has an upper
limit $h_{\mbox{max}} \sim r_p$, where $r_p$ is the radius of polar
cap.  $\Delta V$ is proportionally smaller for longer-period pulsars,
which ultimately results in the sparking process to stop and hence the pulsar
emission to switch off.

\subsection{GM01 model of IVG formation} \label{sec:IVG}
\noindent

GM01 explored the formation of IVG in presence of strong surface
magnetic field and suggested the condition $T_s = T_i$, which can be
expressed as critical lines in the $P-\dot{P}$ diagram.  The polar cap
is heated due to bombardment of backflowing electrons created during
the sparking process and subsequently accelerated by $\Delta V$.  It
can be estimated that $T_s = (\kappa F)^{1/4} \left(e \Delta V
\dot{N}/\sigma \pi r_p^2\right)^{1/4}$, where $\dot{N} = \pi r_p^2
B_s/eP$ is the particle flux through the polar cap, $r_p = (1.4 \times
10^4) b^{-0.5} P^{-0.5}$ cm, and $\kappa F$ is a reduction parameter
introduced by GM01 to accommodate for the change in speed of the
bombarding particles during the spark development process.  As seen
above, $T_s$ depends on $\Delta V$ which is a function of $h$.  The
estimates of $h$ are dependent on models of mean free path of pair
producing photons in the IVG, as well as the spark-formation process.
We briefly discuss below the arguments for obtaining $h$ which are
essential for the results of GM01.

In the RS75 model, the IVG discharges in the form of isolated sparks
at several places by the Sturrock mechanism
\citep{1971ApJ...164..529S}, which is described as follows: When a
photon of energy in excess of 1 MeV is incident in a strong and curved
magnetic field, it can split into an electron-positron pair.  The
large electric field in the gap causes the positrons to accelerate
away from the star, while the electrons are accelerated towards the
surface.  These accelerating charges further radiate curvature
photons; this forms additional pairs along adjacent field lines.  The
process continues until the whole sparking region is screened by
Goldreich-Julian charge density $n_{\mbox{\textsc{gj}}}$.  During this
process, the size of the sparking region grows both in the vertical
($h_\parallel$) and horizontal ($h_\perp$) directions.  RS75 proposed
that $h_\parallel \approx h_\perp$ = $h$.  The gap height is
determined by the mean free path of the accelerating charge particles
($l_{\mbox{part}}$) to produce high energy photons, and the mean free
path of photons ($l_{\mbox{ph}}$) before they are absorbed in magnetic
field to produce pairs, i.e., $h \sim l_{\mbox{part}} +
l_{\mbox{ph}}$.

\begin{figure*}
\begin{center}
\includegraphics[scale=0.55,angle=0.]{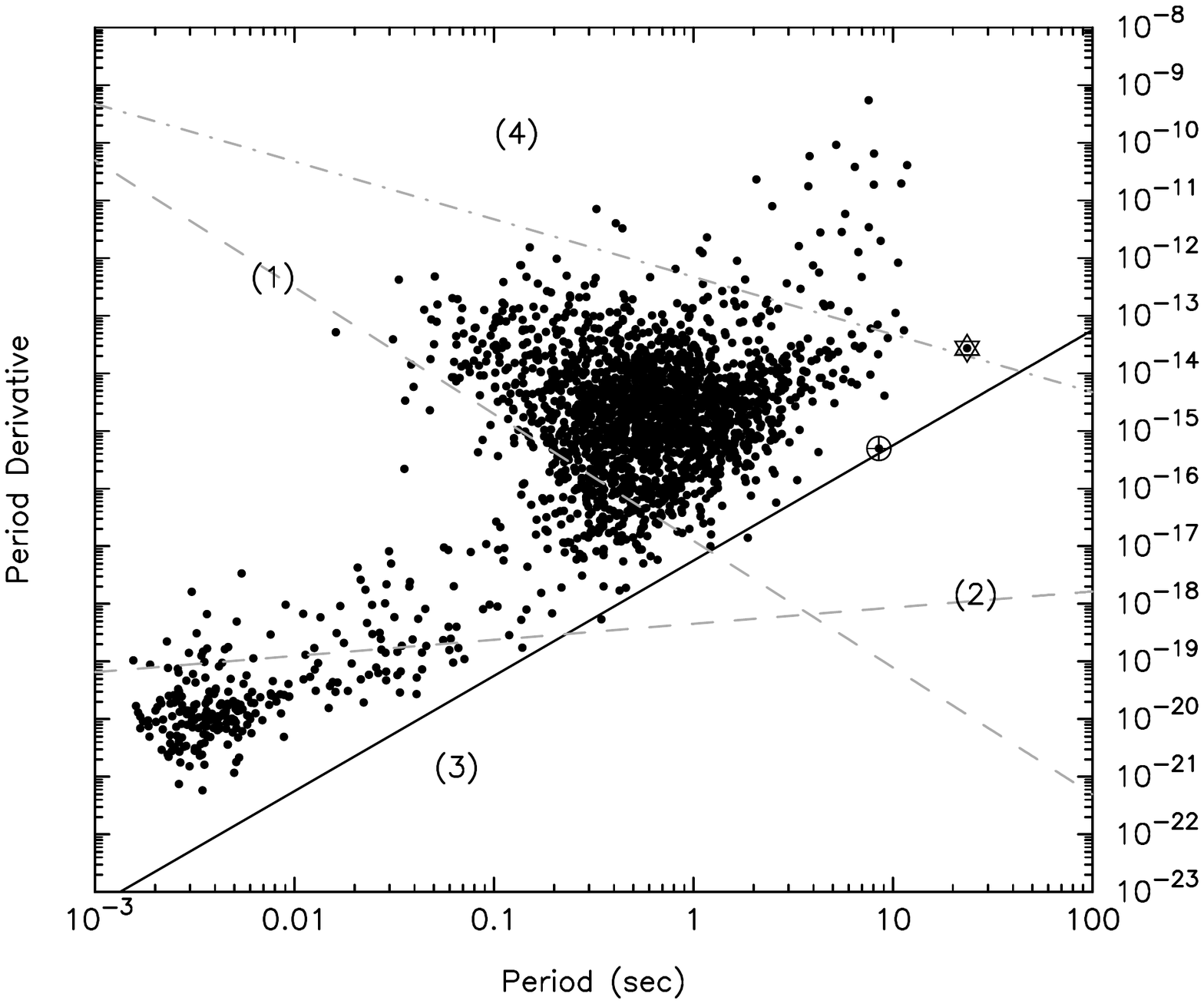}
\caption{\label{fig_ppdot}The $P-\dot{P}$ distribution of all known
  pulsars as obtained from the ATNF catalogue
  \citep{2005AJ....129.1993M}. Lines 1 and 2 in are based on the
  VG-ICS model of GM01 and have been computed using the parameters $b
  = 40$, $\Re_6 = 0.1$ and $\kappa F = 0.5$ (see Sec.\ \ref{sec:IVG}
  for details). Line 1 corresponds to the VG-ICS line for vacuum gap
  formation given by $\dot{P_{-15}} = 2 \times 10^{2} (\kappa F)^{0.7}
  \Re_6^{0.8} b^{-2} P^{-2.2}$. Line 2 represents the VG-ICS death
  line condition $\dot{P}_{-15} = 0.25 \Re_6^{1.14} b^{-1}
  P^{0.28}$. Line 3 is our estimated death line for the PSG model and
  is calculated using Eq.\ \ref{psg3} and parameters $b = 40$, $\Re_6
  = 0.1$, $\eta=0.1$, $T_6 = 2$ and $\alpha_l = 45^{\circ}$. Line 4 is
  obtained from the higher magnetic field limit of the critical
  magnetic field $B_q = m^2c^3/e\hbar = 4.4\times10^{13}$ G beyond
  which the pair creation process is suppressed. The two known
  longest-period pulsars J2144$-$3933 (circle with cross) and
  J0250+5854 (star) have been explicitly identified in the plot.}
\end{center}
\end{figure*}

GM01 noted that the formation of IVG in the normal pulsar population
is possible if $0.1 B_q < B_s \leq B_q$.  In the presence of such
strong magnetic fields, the near-kinematic threshold condition for
pair creation is relevant.  This implies that if a photon with energy
greater than 1 MeV moves in super-strong magnetic fields ($> 0.1
B_q$), then after traversing a distance $l_{\mbox{ph}} \sim 2 m c^2
\Re /\hbar \omega$, pairs at or near the kinematic threshold will be
produced.  Here, $\omega$ is the photon frequency and $\Re$ is the
radius of curvature of the magnetic field.  High-energy photons can be
created in the IVG by accelerating charges to Lorentz factors $\gamma$
via two distinct mechanisms, curvature radiation with characteristic
frequency $\omega = (3/2)\gamma^3 c/\Re$ or resonant inverse Compton
scattering with frequency $\omega = 2 \gamma e B /mc$.  Noting that
$\gamma = e \Delta V/ m c^2$ and using the above relations, the
estimates for $h$ can be obtained in both scenarios.  GM01 applied
these conditions for the two cases:
\begin{enumerate}
 \item Curvature radiation: where $l_{\mbox{part}} << l _{\mbox{ph}}$,
   and hence $h = (3 \times 10^3) \Re_6^{2/7} b^{-4/7} P{-1/7}
   \dot{P}_{-15}^{-2/7}$ cm, here $\Re_6 = \Re/10^6$ and
   $\dot{P}_{-15} = \dot{P}/10^{-15}$.  Thus putting $T_s = T_i$ one
   can find the VG-CR family of critical line as, $\dot{P_{-15}} = 2.7
   \times 10^3 ( \kappa F)^{1.15} \Re_6^{0.64} b^{-2} P^{-2.3}$ (see
   Eq.\ 10 of GM01). The next criteria for pulsar switching off arises 
   from the condition $h_{\mbox{max}} = r_p/\sqrt{2}$, which gives the 
   family of death lines for the VG-CR case as $\dot{P_{-15}} = (2.4 
   \times 10^{-4}) \Re_6 b^{0.5} P^{4.5}$.
 \item Inverse Compton scattering: $l_{\mbox{part}} \sim
   l_{\mbox{ph}}$, and using $l_{\mbox{part}} \sim 0.00276 \gamma^2
   B_{12}^{-1} T_6^{-1}$ as suggested by \citet{2000ApJ...531L.135Z},
   we have $h = (5 \times 10^{3}) \Re_6^{0.57} b^{-1}
   \dot{P}_{-15}^{-0.5}$ cm. The VG-ICS family of critical lines for 
   vacuum gap formation using $T_s = T_i$ is given as, $\dot{P_{-15}} 
   = 2 \times 10^{2} (\kappa F)^{0.7} \Re_6^{0.8} b^{-2} P^{-2.2}$ 
   (see Eq.\ 15 of GM01). The death lines for the VG-ICS case, once 
   again using $h_{\mbox{max}} = r_p/\sqrt{2}$, is estimated as 
   $\dot{P_{-15}} = 0.25 \Re_6^{1.14} b^{-1} P^{0.28}$.
\end{enumerate}
The region bounded by the line of IVG formation and death line in the 
$P-\dot{P}$ diagram corresponds to radio-loud pulsars.  Consequently, 
GM01 argued that IVG can form in majority of the normal pulsars only 
if the surface fields are strong and non-dipolar with $b \sim 50 - 
100$.  They also showed that the VG-ICS mechanism allows more 
pulsars to form an IVG. Fig.\ \ref{fig_ppdot} shows the $P-\dot{P}$ 
distribution of all known pulsars, where line 1 represents the VG-ICS 
boundary for gap formation and line 2 corresponds to the ICS death 
line condition as presented above (and Eq 15 and 16 of GM01).  The 
lines use representative values of the different parameters with 
$b = 40$, $\Re_6 = 0.1$ and $\kappa F = 0.5$.

The spark formation in the RS75 model of IVG has several limitations
as noted by \citet{1977ApJ...214..598C}.  Firstly, the Sturrock
mechanism is difficult to realize in simple magnetic field geometries.
A discharge of electron-positron pair can start at any particular
magnetic field line and then progress in a certain direction without
ever occurring in the original location \citep[see Fig.\ 1
  of][]{1977ApJ...214..598C}.  In the absence of repeating pairs along
a particular location, it is not possible for the initial discharge to
grow into a spark and locally reach Goldreich-Julian density.
\citet{1977ApJ...214..598C} proposed a mechanism called photon splash,
where a back streaming particle penetrates the stellar surface and
causes a high energy ($>$1MeV) $\gamma$-ray to be emitted back in the
gap region.  This photon can now traverse in any direction and even
produce pairs in the original field line: It is difficult to know if
such a process is at work.  The second limitation concerns the
stability of such sparks.  The development of each spark spans
timescales of several tens of microseconds.  On the other hand, the
observed subpulses, which are believed to be associated with the
spark-associated plasma columns, are seen typically for several
milliseconds.  This calls for a process that makes the sparks remain
as stable entities for longer timescale: This is not addressed in
RS75.  Finally, as discussed in GMG03, the expected subpulse drift
rates from the IVG model of RS75 is much faster that the observed
values and, similarly, the estimated surface temperatures from this
model are significantly higher than those measured from the polar cap.

\subsection{PSG model and application to PSR J2144$-$3933}
\noindent

The possibility of steady outflow of ions from the surface of the
neutron star was first suggested by
\citet{1977ApJ...214..598C,1980ApJ...235..576C}.  This concept has
been used by GMG03 to further develop the partially screened gap (PSG)
model of the inner acceleration region.  \citet{1980ApJ...235..576C}
suggested that even if the surface of the neutron star is below $T_i$,
there are still ions that can flow from the surface.  Hence, the IVG
will not exist in a purely vacuum state but will be partially filled
with ions.  In the PSG model, the ion charge density ($\rho_i$) is
related to $\rho_{GJ}$ as $\rho_i/\rho_{GJ} \approx \exp(30 -
T_i/T_s)$, where $\rho_i \leq \rho_{GJ}$ \citep[see Eq.\ 8 in][and
  Eq.\ 1 in GMG03]{1980ApJ...235..576C}.  If the charge density is
lower than $\rho_{GJ}$ the accelerating potential in the inner
accelerating region is reduced by a factor of $\eta$ such that $\Delta
V_{\mbox{\textsc{psg}}} = \eta \Delta V$, where $\eta = 1 -
\rho_i/\rho_{GJ}$.  The exponential dependence of charge density on
$T_s$ also makes $\Delta V_{\mbox{\textsc{psg}}}$ very sensitive to
its variations; i.e., extremely small changes in $T_s$ can lead to
large changes in the gap potential.  As a result, it was proposed that
the PSG is maintained in perfect thermal equilibrium with $T_s$ being
close to $T_i$ \citep[][GMG03]{1980ApJ...235..576C}.  The thermal
regulation can be visualized in the following manner: As soon as the
temperature at a given region in the polar cap goes slightly below
$T_i$, the accelerating potential increases and pair creation can take
place which instigates the sparking process.  As the spark grows,
$T_s$ increases and hence $\eta$ continues to increase until $T_s =
T_i$.  At this stage, the spark terminates and the heating stops
causing the surface to cool such that $T_s < T_i$, and the process
repeats itself.

Let us now examine the conditions in the PSG model for the formation of a spark.
The potential drop across a spark with width $h_{\perp}$ is \citep{2013arXiv1304.4203S}
\begin{equation}
\Delta V_{\mbox{\textsc{psg}}} = \frac{4 \eta \pi b B_d \cos{\alpha}_l}{c P} h_{\perp}^2 =
\frac{4 \eta b B_d \cos{\alpha}_l}{cP} A_{\mbox{sp}}.
\label{psg1}
\end{equation}
Here, $A_{\mbox{sp}} = \pi h_{\perp}^2$ is the area of the spark and
$\alpha_l = \alpha + \mu$ is the angle between the local magnetic
field and the rotation axis, $\alpha$ is the angle between the
rotation axis and the dipolar magnetic field component (see
Sec.\ \ref{sec_geom}), and $\mu$ quantifies the deviation from dipole
behaviour.  A detailed discussion of the underlying assumptions of
Eq.\ \ref{psg1} is presented in Appendix \ref{as1} (see also the
discussion above Eq.\ \ref{a13}).  As a result of the partial
screening of the gap by factor $\eta$, the number of positrons
required by the sparking process to reach Goldreich-Julian density is
reduced to $\eta n_{\mbox{\textsc{gj}}} = 1.4 \times 10^{11} \eta
(\dot{P}_{-15}/P)^{0.5} b \cos{\alpha}_l$.  An equal number of
electrons moves in the opposite direction towards the polar cap
surface and supplies an energy equivalent of $L_{\mbox{part}} = q
\Delta V_{\mbox{\mbox{\textsc{psg}}}} \eta n_{\mbox{\textsc{gj}}} c =
\eta^2 (\dot{P}_{-15}/P^2) (A_{\mbox{sp}}/A_{\mbox{pc}}) b
\cos^2{\alpha_l} $, where $A_{\mbox{pc}} = 6.58 \times 10^{8}
(bP)^{-1}$ cm$^2$.  The condition for thermostatic equilibrium in the
gap, $L_{\mbox{part}} = \sigma T^4$, leads to the relation
\begin{equation}
\frac{A_{\mbox{sp}}}{A_{\mbox{pc}}} = 3.16 \times 10^{-4} \frac{T_6^4}{\eta^2 b \cos^2{\alpha_l} } \frac{P^2}{\dot{P}_{-15}},
\label{psg2}
\end{equation}
where $T_6$ = $T$/10$^6$ K.
The sparks can only form when $A_{\mbox{sp}}\leq A_{\mbox{pc}}$.
In the limiting case when $A_{\mbox{sp}}= A_{\mbox{pc}}$, the death line for the PSG model is
\begin{equation}
\dot{P}_{-15} = 3.16 \times 10^{-4} \frac{T_6^4}{\eta^2 b \cos^2{\alpha_l} } {P^2}.
\label{psg3}
\end{equation}
This condition for the death line obtained from the PSG model depends
on a number of parameters such as $\eta$, $b$, $\alpha_l$ and $T_6$
which are not well constrained either from observations or from
modeling.  However, in contrast to the IVG model, photon and particle
mean free paths used to estimate the gap heights and widths are no
longer necessary. The IVG model requires two distinct conditions for
the generation of coherent radio emission (criteria for vacuum gap
formation and criteria for sparking discharge) leading to two
different expressions (see Sec.\ \ref{sec:IVG}).  In the PSG model,
both these conditions are captured in the single expression of
Eq.\ \ref{psg3}.

We now investigate the conditions that will enable the single spark
model (Eq.\ \ref{psg3}) to be applied to PSR J2144$-$3933.  This
requires making reasonable estimates for the different physical
parameters in Eq.\ \ref{psg3}.  The polar cap temperature and actual
polar cap area can be obtained from X-ray observations modeled as
blackbody spectra.  PSR J2144$-$3933 is an old pulsar with
characteristic age $2.7 \times 10^{8}$ years and is expected to be
significantly cooler than younger stars.  It is possible to detect the
blackbody emission from the whole star using optical observations.
However, deep X-ray and optical observations have not detected any
significant emission from this pulsar
\citep{2011MNRAS.412L..73T,2019ApJ...874..175G}.  The upper limit for
the neutron star temperature from optical observation is $4\times10^4$
K \citep{2019ApJ...874..175G}, while the hot polar cap temperature is
lower than $1.9\times 10^{6}$ K assuming a radius of 10 m
\citep{2011MNRAS.412L..73T}.  The polar cap size requires the
non-dipolar surface magnetic field to be around hundred times stronger
than the dipolar value; i.e., $b\sim$ 100; and the corresponding ionic
temperature is $T_i \approx 5 \times 10^{6}$ K
\citep{1986MNRAS.218..477J}.  A somewhat more refined estimate of
$T_i$ as a function of $b$ is $T_i = 1.6 \times 10^4
\left[\left(b\sqrt{P\dot{P}_{-15}}\right)^{1.07}+ 17.68\right]$ K
\citep{2013arXiv1304.4203S,2015MNRAS.447.2295S,2007MNRAS.382.1833M}.
For PSR J2144$-$3933 parameters with $b$ = 100 once again leads to
$T_i \approx 5\times 10^{6}$ K.  There is no accurate way to calculate
$T_i$ given the unconstrained nature of $b$.  However, from the upper
limits obtained above we expect $b\sim 40$.  Correspondingly,
$T_i/10^6 = T_6 \approx 2 $ K appears to be a reasonable approximation
for PSR J2144$-$3933.  Next, we concentrate on constraining the
screening factor in the gap.  It has been shown that when $T_i \sim
1-2 \times 10^6$ K and $\Re_6 \sim 0.1-1$, then a wide range of $\eta$
between 0.1 and 0.2 is possible (see Sec.\ 3 and Fig.\ 1 in GMG03).
Additionally, the subpulse drifting velocity in the PSG model is
dependent on $\eta$ (see Eq.\ \ref{a12} in Appendix \ref{as1}).  GMG03
and \citet{2013arXiv1304.4203S} used the observed drift rates for a
number of pulsars and found $\eta$ to be around 0.1.  Another
unconstrained parameter in Eq.\ \ref{psg3} is $\alpha_l$.  In section
\ref{sec_geom} we estimated $\alpha \sim 72^{\circ}$ from polarization
measurements.  However, the non-dipolar field on the surface can
significantly change $\alpha_l$ from the dipolar value by several tens
of degrees \citep{2002A&A...388..235G}.

We have used representative values of $\eta=0.15$,
$\alpha_l=45^{\circ}$, $T_6=2$ and $b=40$ to calculate the death line
of Eq.\ \ref{psg3}, which is shown in the $P-\dot{P}$ diagram of
Fig.~\ref{fig_ppdot} (line 3 in the plot).  The pulsar J2144$-$3933
has been explicitly identified in the plot (circle with cross) and is
roughly coincident with the death line.  This suggests that the actual
polar cap size of this pulsar is such that only one fully grown spark
can form.  The figure also shows the longest period ($P$ = 23.5 sec)
pulsar J0250+5854 (the second star-shaped point in
Fig.\ \ref{fig_ppdot}) which is seen to be further away from the death
line.  For comparison, we have estimated $A_{\mbox{pc}}/A_{\mbox{sp}}$
from Eq.\ \ref{psg2} using the same parameters as above and
$\dot{P}_{-15}$ = 27.2 for this pulsar.  The ratio
$A_{\mbox{pc}}/A_{\mbox{sp}} \approx 5$, which suggests that several
sparks can be formed in its polar cap and is consistent with its
location away from the death line.  The spark associated plasma
columns results in subpulses, and it is expected that more than one
component should be seen in the profile of PSR J0250+5854 (subject to
line-of-sight geometry).  The observed profile at 350 MHz shows the
presence of two components \citep[Fig.\ 6 in][]{2018ApJ...866...54T},
which is consistent with the predictions of the PSG model.  Detailed
polarization and single-pulse studies of this pulsar should provide
more insights into the sparking process.

The death line in Fig.~\ref{fig_ppdot} (line 3) is model-dependent;
i.e., each pulsar will have slightly different value of $\eta$, $B_s$,
$\alpha_l$, etc.  However, the known pulsar population
(Fig.\ \ref{fig_ppdot}) lies above the death line predicted by the PSG
model.  In the higher magnetic field limit, beyond the critical
magnetic field $B_{q}$ (line 4 in Fig.\ \ref{fig_ppdot}), pair
creation processes are expected to be suppressed, thereby terminating
the radio emission \citep[see ][]{1995AuJPh..48..571U,
  1998ApJ...507L..55B,2014ApJ...784...59S}.  Hence, very few radio
pulsars should be seen above line 4.  Most of the points in
Fig.\ \ref{fig_ppdot} above this limit correspond to anomalous X-ray
pulsars which do not emit in radio frequencies.  However, a few radio
pulsars are still detected above this range \citep[see
  e.g.][]{2000ApJ...541..367C} which cannot be explained from the
present theory.  There is also a potential selection bias that may
result in non-detection of pulsars above line 4 \citep[such as narrow
profiles;][]{1999Natur.400..848Y}.  Line 4 is, therefore, more
illustrative than a rigid boundary.  In summary, the PSG model
constrains the radio-loud pulsar population to lie in the region
between lines 3 and 4 (Fig.\ \ref{fig_ppdot}) in the $P-\dot{P}$
diagram and, by and large, this prediction is validated by the
currently known pulsar population.

\subsection{Understanding the radio emission properties of PSR J2144$-$3933} 

We have shown that the PSG model, under reasonable assumptions,
predicts a single spark operating in PSR J2144$-$3933.  In the
following discussion, we have used this hypothesis to understand the
properties of observed radio emission described in
Sec.\ \ref{sec_obs2}.

\paragraph*{Subpulse Drifting.}
Our detailed analysis in section \ref{sec_Imod} indicated that PSR
J2144$-$3933 shows no subpulse drifting. Here, we explore the origin
of subpulse drifting within the framework of the PSG model and
investigate the conditions under which it will vanish. When the
temperature of the surface is slightly greater than $T_i$, the ions
can flow freely from them. Random fluctuations in temperature can lead
to conditions where $T_s$ goes below $T_i$ in certain regions,
inhibiting the ion flow above it. A high potential drop quickly
develops above this region whose form is given by Eq.\ \ref{psg1}.
Consequently, sparking discharge is induced which continues to grow
till the spark width reaches $h_{\perp}$. Typically, this takes
several microseconds (see e.g. \citealt{2000ApJ...541..351G},
  GMG03) during which the region beneath the spark is heated due to
bombardment of backflowing electrons accelerated in the gap. When the
charge density in the spark reaches the Goldreich-Julian density
$n_{\mbox{\textsc{gj}}}$, the temperature on the surface $T_s$ also
increases to slightly above $T_i$, and the sparking process
terminates. At this stage the plasma column in the gap co-rotates with
the star till the gap is emptied. In this entire sparking process, the
hot spot on the surface is left slightly lagging behind the
co-rotation velocity. The gap emptying time is also expected to be of
the order of microseconds. In contrast, the hot spot cooling time is
nanoseconds (for timescale estimates see GMG03) which is several
orders of magnitude lower. Since the temperature at the lagging hot
spot region drops faster, the condition $T_s < T_i$ is satisfied and
the region cools down, and the sparking process starts at the same
lagged behind region. This lagging behind co-rotation of the hot spot
region, its subsequent cooling down and regeneration of the sparking
process also ensures the spark associated plasma to lag behind
co-rotation as emphasized in earlier works of
\citet{2016ApJ...833...29B,2018MNRAS.475.5098B,
  2019MNRAS.482.3757B}. The growth of plasma instabilities in the
non-stationary spark-associated plasma flow results in coherent radio
emission higher up in the magnetosphere. As a result, the emission
also lags behind co-rotation speed and is observed as subpulse
drifting.

In Appendix \ref{as1}, we describe in detail the general drifting
behaviour in an inclined rotator using the PSG model. We see clearly
that during the time in which the spark develops and attains
$n_{\mbox{\textsc{gj}}}$, the plasma is lagging behind the co-rotation
of the neutron star (Eq.\ \ref{adrift}). It is possible to measure the
drifting periodicity ($P_3$) which corresponds to the time taken by
the sparks to repeat at any pulse phase
\citep{2013arXiv1304.4203S,2016ApJ...833...29B}. A detailed study of
the drifting behaviour was carried out by \citet{2016ApJ...833...29B},
where using predictions from the PSG model a dependence of $P_3$ on
observable parameters was obtained. It was shown that
$P_3^{\mbox{\textsc{psg}}} \approx 2 (\gamma_6/\xi_{-3})(\dot{E}/(4
\times 10^{31}))^{-0.5} P$, where $\gamma_6 \sim \gamma/10^6$ is the
Lorentz factor of the primary particles formed in sparks, $\xi_{-3} =
\xi/10^{-3}$ can be related to the efficiency of non-thermal emission
and obtained from X-ray observations, and $\dot{E}$ is the spin-down
energy loss expressed in erg s$^{-1}$. The $P_3-\dot{E}$ dependence is
an important relationship because it can be verified through
observations.  However, measurements of $P_3$ are fundamentally
uncertain due to the aliasing effect, where observationally it is
impossible to differentiate between $P_3$ ($>2P$) and its aliased
value $P_3^{'}= P_3/(P_3$-1) ($P < P_3^{'} < 2P$). A physical
  model is essential to disentangle $P_3$ and $P_3^{'}$, and the
  lagging behind co-rotation speed for the subpulses in PSG model
  allows a resolution to the aliasing effect. The lagging subpulses
are expected to move from the leading to the trailing edge of the
pulse window. This implies that negative drifting, where subpulses
shift towards the leading part of the profile in subsequent periods,
has $P_3 > 2P$, while positive drifting, with subpulses shifting
towards the trailing edge, has $P < P_3 < 2P$. A comprehensive
observational campaign has been conducted to measure drifting in the
pulsar population by
\citet{2016ApJ...833...29B,2019MNRAS.482.3757B}. It was found that
drifting is limited to pulsars with $\dot{E} < 2.3\times10^{32}$
erg~s$^{-1}$. The measured $P_3$, using the above model, has a certain
spread but is anti-correlated with $\dot{E}$, given as
$P_3^{\mbox{\textsc{obs}}} \approx (\dot{E}/(2.3\times
10^{32}))^{-0.6} P$. This shows that the PSG model of subpulse
drifting is consistent with observations.

The pulsar J2144$-$3933 has $\dot{E}$ = 3.18$\times10^{28}$
erg~s$^{-1}$, and is expected to show negative drifting with subpulses
moving from the trailing to the leading edge of the profile at
subsequent periods.The expected periodicity of drifting from the PSG
model is $P_3^{\mbox{\textsc{psg}}} \sim 70P$, and that estimated from
the observational fits is $P_3^{\mbox{\textsc{obs}}} \sim 205P$. The
variations between the two can be explained from the scatter in the
observational measurements of $P_3$ as well as the uncertainty in
estimating the parameter $\xi$
\citep{2012ApJS..201...37K,2016ApJ...833...59S}. As reported in
section \ref{sec_Imod} we have not detected the presence of
drifting. There is a possibility that the drifting maybe hidden by the
scintillation timescale ($\geq 400P$), but this is relevant only in
case of most extreme estimates of $P_3$. We suggest that the absence
of subpulse drifting in PSR J2144$-$3933 is due to the presence of a
single spark in a small polar cap with limited space.  If the polar
cap is significantly larger than the spark size, several sparks can
form in a closely packed manner, where the whole pattern lags behind
co-rotation motion. In these cases, there is sufficient space for the
sparks to continuously form as well as move across the polar cap to
exhibit subpulse drifting. However, for PSR J2144$-$3933, once the
spark is fully formed, it fills the whole polar cap and there is no
further space for it to move across.  The spark terminates and is
regenerated at the same place. We further conjecture that lack of
drifting in certain pulsars can arise due to conditions in the polar
cap which inhibit sparks to move across it.

\paragraph*{Nulling.}
In Sec.\ \ref{sec_Imod}, we found PSR J2144$-$3933 predominantly shows
single-period nulls. The pulse window, in this case, is approximately
2\degr~wide in longitude, corresponding to roughly 50 msec in time,
which is the minimum time duration for the radio emission to switch
off during nulling.  The presence of one single spark in this pulsar
also rules out the possibility that the short nulls results from the
line of sight passing between empty regions of the sparking system
\citep[see e.g.][]{2009MNRAS.393.1391H}. The empty line of sight
argument for short-duration nulls have been considerably weakened by
exhaustive studies showing their presence in the core component
\citep{2017ApJ...846..109B,2018MNRAS.475.5098B,2019MNRAS.486.5216B}. This
leaves two extreme possibilities: Either (a) the plasma flow entirely
stops, or (b) the plasma flow is maintained but the conditions for
coherent radio emission break down. According to the PSG model, the
first case (a) corresponds to $T_s < T_i$, and it is difficult to
envisage this condition to hold for several milliseconds. Even if it
is possible to have $T_s < T_i$ for millisecond durations, the
high-energy gamma-ray photons from the diffuse background can easily
create pairs in the region above the polar cap within a few
microseconds \citep[see][]{1982ApJ...258..121S}. Hence, it is highly
unlikely that the plasma flow is stopped for the duration of the
nulls. The second alternative (b) implies that the thermostatic
regulation of the PSG is maintained, but the nature of the
non-stationary plasma flow is modified. This requires the
pair-creation properties to change: In principle, this allows certain
possibilities such as short-term change in the structure of the
non-dipolar surface magnetic field, etc. However, at present, we are
unaware of any model that can explain such changes at timescales of
milliseconds to seconds.
\begin{figure}
\begin{center}
\includegraphics[scale=0.38]{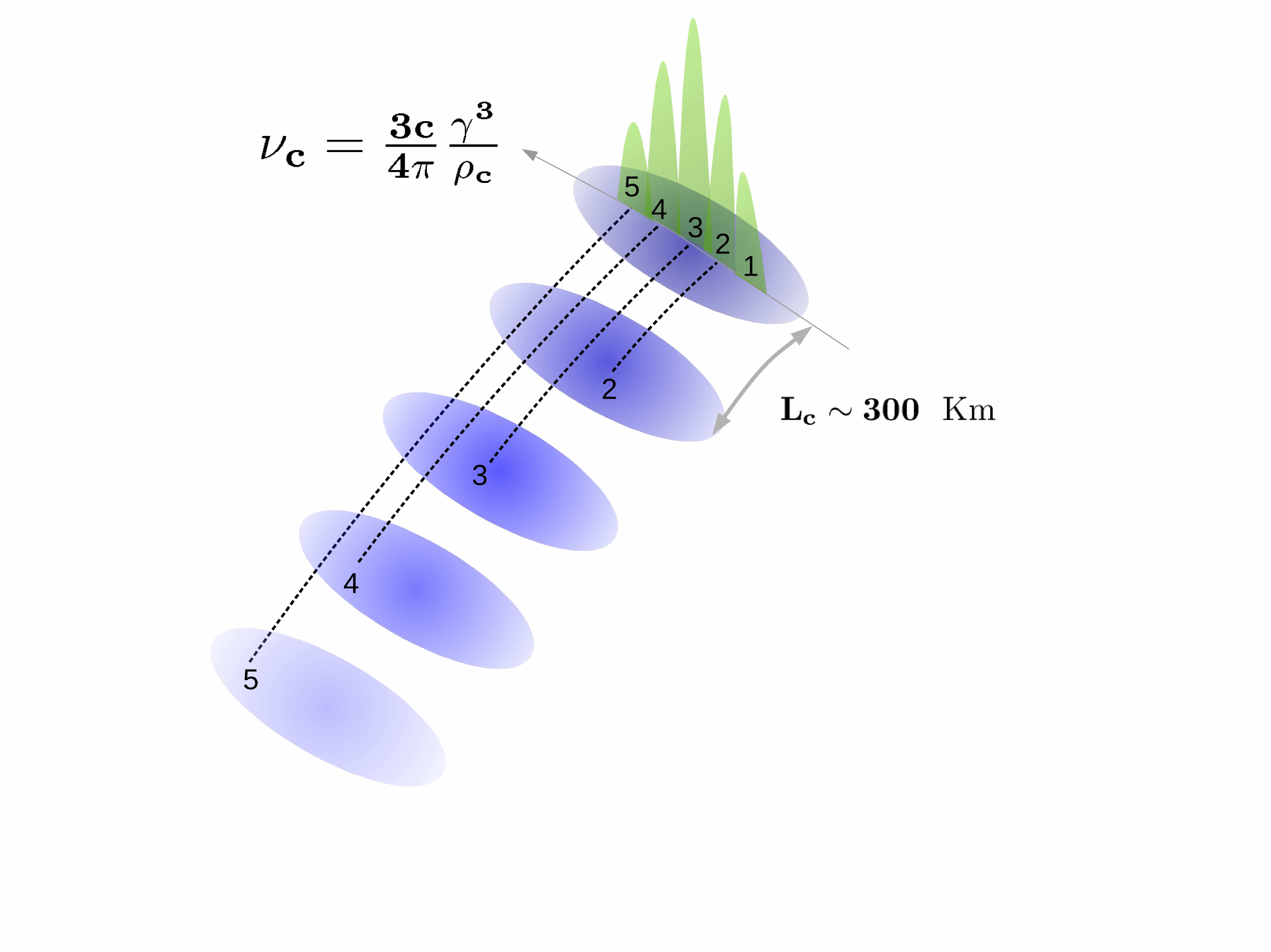}
\caption{\label{fig_schem}A cartoon illustration of how the
  quasiperiodic structures arise in subpulses. The blue regions
  correspond to five strong emission sources which are moving
  relativistically along the open field lines.  The separation between
  adjacent regions as well as their physical extent is given by $L_c =
  c \Delta t$. For the typical value $\Delta t \sim 1$ msec, we get
  $L_c \sim 300$ km. Radio emission originates from a specific height
  due to curvature radiation with a characteristic frequency
  ($\nu_c$). The five regions identified in the figure cross this
  emission height sequentially, separated by $\Delta t$ in time, are
  detected in the emission window as separate peaks, and the empty
  regions between them are seen as dips in the intensity pattern. Due
  to the rotation of the pulsar, each emission source is observed at
  delayed phases within the pulse window. The linear relation of
  modulation periodicity $P_{\mu}$ with pulsar period $P$ suggests
  that the length scale $L_c$ remains constant for different pulsars.}
\label{fig_schem}
\end{center}
\end{figure}

\paragraph*{Quasiperiodic structures.}
As seen in Fig.~\ref{fig_micro}, PSR J2144$-$3933 has the presence of
modulating, quasiperiodic structures in its subpulses. This effect is
seen in several pulsars and, when such modulating structures are
clearly resolved, the periodicity of modulation $P_{\mu}$ is
proportional to the pulsar period $P$ as $P_{\mu} \approx 10^{-3} P$
as shown in \citet{2015ApJ...806..236M}. This study suggested that
such radiation patterns can result from the temporal modulations of
non-stationary plasma flow, associated with the sparking process in
the vacuum-gap class of models.  Coherent radio emission is believed
to be excited due to curvature radiation from charge bunches at
regions below 10\% of the light cylinder
\citep{2004ApJ...600..872G,2009ApJ...696L.141M,2017JApA...38...52M,
  2018MNRAS.480.4526L}. The characteristic frequency of emission is
$\nu_c = (3/4\pi) \gamma^3c/\rho_c$, where $\gamma$ is the Lorentz
factor of the charge bunch and $\rho_c$ is the radius of curvature of
the field lines. Since $\gamma$ does not change significantly and the
emission arises from regions of dipolar magnetic field lines, radio
emission at any given frequency arises from a narrow range of
heights. The quasiperiodic structure seen across any subpulse should
also arise from this narrow region. The origin of these
microstructures can be understood under the assumption that there is a
constant stream of modulating radiation patterns with lateral angular
dimension equivalent to the subpulse width moving relativistically
along the open magnetic field flux tubes. The radial lengths of these
structures are a few hundred kilometers which are separated from each
other by distinct empty regions. As the pulsar rotates, our line of
sight encounters these alternating emitting structures followed by the
relatively empty spaces between them at different phases of the
subpulse, resulting in the observed modulating emission pattern. In
Fig.\ \ref{fig_schem}, we present a schematic diagram to aid the
visualization of a steady stream of emitting regions observed as
microstructures. The period dependence of the modulation periodicity
can be explained if we assume that the radial dimensions of these
emitting regions remain unchanged across the pulsar population.

The single-spark PSG model for PSR J2144$-$3933 shows the
  sparking to be a result of the single stream of outflowing plasma,
  which provides a simplified window into its IAR. According to RS75,
the gap height in the IAR is around 100 meters. The sparking process
in the gap region lasts for a few tens of microseconds during which
the $n_{\mbox{\textsc{gj}}}$ charge density is attained.
Subsequently, the gap empties within about half a microsecond
resulting in a spark-associated plasma column. The radial dimension of
this plasma cloud is around three kilometers and it moves
relativistically outward from the stellar surface along the open field
lines. The separation between adjacent clouds corresponding to the
gap-emptying time is around hundred meters. In steady state, just
above the gap region, a non-stationary flow of plasma is
established. Each plasma cloud has a distribution of particle speeds.
During their outward motion, the low and high velocities of the clouds
overlap at a certain height, leading to a two-stream instability and
the growth of plasma waves. For strong growth of plasma waves the
system can be driven to nonlinear regimes, where charge bunches can
excite coherent curvature radiation. Depending on the growth and decay
of plasma waves, bright and weak radio emission patterns can be
generated. However, the timescales associated with these patterns are
around several microseconds, corresponding to the plasma cloud sizes,
and cannot explain the typical millisecond variations associated with
the quasiperiodic structures. Additional physical processes
  resulting in variations of the sparking processs over timescales of
  milliseconds are required to explain the observed quasiperiodic
  structures which are currently unknown. Our estimates require such
  variations to span over typical length scales of 300 km, as
  illustrated in Fig.\ \ref{fig_schem}.

In the PSG model, the non-stationary flow is expected to be
established in a manner similar to the prescription of RS75. However,
the screening potential will modify the gap height and consequently
the length of the instability region and the cloud overlap
timescales. A detailed study of the spark development process in the
PSG combined with the growth of plasma instability in the radio
emission region is needed to get estimates of the cloud size and
instability region timescales. Such estimates is
  currently unavailable, however it is still unlikely that the
modified timescales from the PSG model will be sufficient to explain
the millisecond quasiperiodic structures and additional physical
processes will still be needed.

\section{Summary} 
We have estimated a new death line in the $P - \dot{P}$ diagram of the
pulsar population using the single-spark PSG model
(Eq.\ \ref{psg3}). Our death line also explains the scarcity of
long-period pulsars, contradicting the earlier predictions of
GM01. Specifically, we show that this model is able to explain the
radio-loudness of PSR J2144$-$3944 as being located very close to this
boundary, and that the radio emission properties of PSR J2144$-$3944
are consistent with this model. Our study provides strong support for
theoretical models which predict an inner acceleration region in radio
pulsars where the sparking process just above the polar cap results in
a nonstationary relativistic plasma flow essential for the generation
of coherent radio emission.

\section*{Acknowledgments} \noindent We thank the referee for comments which 
improved the paper. We thank the staff of the GMRT who have made these
observations possible. The GMRT is run by the National Centre for
Radio Astrophysics of the Tata Institute of Fundamental Research. DM
acknowledges support and funding from the `Indo-French Centre for the
Promotion of Advanced Research – CEFIPRA' grant IFC/F5904-B/2018.

\appendix 

\section{Golreich Julian Density and Subpulse drift direction and Speed in 
PSG model} 
\label{as1}
In a slowly rotating neutron star system we consider two frames of
reference, primed and unprimed, where the primed correspond to the
co-rotating frame located at the neutron star surface, and the
unprimed to the observer frame.  For slow rotation, the velocity
  at the neutron star surface is significantly less than the velocity
  of light and hence the two frames can be considered as inertial
  frames. In this case the components of the electric field between the
  two frames is transformed as

\begin{equation} 
\vec{E^{\prime}} = \vec{E} + \frac{1}{c} \left( \vec{\Omega} \times \vec{r} \right) \times \vec{B} 
\label{a1} 
\end{equation} 
Here $\vec{\Omega}$ is the rotation vector, $\vec{r}$ is the radial
vector and $\vec{B}$ is the magnetic field.  Since above the polar cap
the quantity $\Omega r / c$ is negligible, then the term $\delta
B/\delta t \sim 0$, and thus the maxwell equations in the gap region
are mainly guided by electrostatics. If the charge density is given by
$\rho(r,\theta,\phi)$ then the Gauss law and Faraday's law
respectively can be written as,

\begin{equation} 
\mathbf{div}{\vec{E^{\prime}}}
=  4\pi \rho(r,\theta,\phi) + \div{ \left(
\frac{1}{c} (\vec{\Omega} \times {\vec{r}} ) \times \vec{B}\right)} 
\label{a2}
\end{equation}

\begin{equation}
\mathbf{curl}{\vec{E^{\prime}}} = 0
\label{a3}
\end{equation}

Applying appropriate boundary condition the electric field
${\vec{E^{\prime}}}=0$, when $\rho(r,\theta,\phi) = \rho_{GJ} =
-1/(4\pi)\div{ \left(\frac{1}{c} (\vec{\Omega} \times {\vec{r}} )
  \times \vec{B}\right)}$.  Further if one assumes that in the polar
cap region $B_{\theta} = B_{\phi} = 0$ and $B_r \neq 0$, which is
generally a very good assumption for both dipolar and non-dipolar
magnetic fields, then one gets the usual form of the Goldreich Julian
density as,

\begin{equation}
\rho_{GJ} = - \frac{\vec{\Omega} \cdot \vec{B}}{2 \pi c}
\label{a4}
\end{equation}  

Note in Eq.~~\ref{a2} the term in the left hand side can be simplified
as $((\vec{\Omega} \times {\vec{r}} ) \times \vec{B}) =
\vec{r}\cdot(\vec{\Omega}\cdot\vec{B}) -
\vec{\Omega}\cdot(\vec{r}\cdot\vec{B})$. For the case $B_{\theta} =
B_{\phi} = 0$ and $B_r \neq 0$, this becomes $((\vec{\Omega} \times
{\vec{r}} ) \times \vec{B}) = (r \Omega B_r \sin\theta) \vec{\theta}$,
here $\vec{\theta}$ is the unit vector.  Now applying the divergence
theorem to Eq.~~\ref{a2} one gets,

\begin{equation}
\iint \limits_{S}^{}{ {\vec{E^{\prime}}}} dS=  4\pi \iiint \limits_{V}^{} \rho(r,\theta,\phi) dV+ 
\frac{1}{c}\iint\limits_{S}^{} (r \Omega B_r \sin\theta\cdot \hat{\theta})  dS
\label{a5}
\end{equation}

Now replacing $\rho = \rho_{GJ} - \Delta \rho$, and rearranging the
equation one can write,

\begin{equation}
\iint \limits_{S}^{}({ {\vec{E^{\prime}}}}-\frac{1}{c} (r \Omega B_r \sin\theta\cdot \hat{\theta}))  dS
=  4\pi \iiint \limits_{V}^{} (\rho_{GJ} - \Delta \rho) dV 
\label{a6}
\end{equation}

We can now write $\vec{{E^{\prime}}} = E^{\prime}_r \hat{r} +
E^{\prime}_{\theta} \hat{\theta} + E^{\prime}_\phi\hat{\phi}$. Let us
now consider volume element such that one side touches the wall of
vacuum gap and also note that $E^{\prime}_\phi = 0$. We also make the
reasonable assumption that $E^{\prime}_r$ does not change
significantly across the gap which gives $\iint \limits_{S}^{}
E^{\prime}_r ds = 0$. Thus Eq.~\ref{a6} can now be written as,

\begin{equation}
\iint \limits_{S}^{} ({ {{E^{\prime}_{\theta}\hat{\theta}}}}-
\frac{1}{c} (r \Omega B_r \sin\theta\cdot \hat{\theta}))  dS
=  4\pi \iiint \limits_{V}^{} (\rho_{GJ} - \Delta \rho) dV
\label{a7}
\end{equation}

We can now consider two extreme cases, one where $\rho = \rho_{GJ}$
such that $\Delta \rho = 0$. In this case as per Eq.~\ref{a7} ,
$E^{\prime}_{\theta} = 0$, and this means that the entire electric
field in the co-rotating frame is screened and this correspond to the
case of perfect co-rotation of the charges in the IVG. Next we
consider the case where $\rho = 0$ or $\Delta \rho = \rho_{GJ}$. In
this case, $E^{\prime}_{\theta} = (1/c) r \Omega B_r \sin\theta$,
which means that there is full electric field available in the IVG,
and hence a test charge will appear to be lag behind the rotation in
the co-rotating frame and remain at rest in the observer frame. For
any intermediate case,

\begin{equation}
\iint \limits_{S}^{}{ {{E^{\prime}_{\theta}\hat{\theta}}}} dS
=  -4\pi \iiint \limits_{V}^{} \Delta \rho dV
\label{a8}
\end{equation}

$E^{\prime}_{\theta}$ is directed in such a direction that it will
always be directed behind the linear velocity $v_d^{\prime}$, as
indicated by the negative sign i.e.
\begin{equation}
\vec{v_d^{\prime}} = (\vec{E}^{\prime} \times \vec{B}^{\prime}/{B^{\prime}}^2) c = - c (E^{\prime}_\theta/B_r) \hat{\phi}
\label{adrift}
\end{equation}
 Here $v_d^{\prime}$ is also known as the drift velocity and to
  calculate the magnitude of {\bf $v_d^{\prime}$} in terms of gap
  parameters we need to estimate $E^{\prime}_{\theta}$.


In RS75 (see their Appendix and Eq. A7 b,c ) the IVG accelerating
potential for an anti-aligned rotator was calculated as, $\Delta V =
(\Omega B/c) h^2$, where $h$ corresponds to the height of the
gap. CR80 showed that the stable gap potential solutions can also
exist for an inclined rotator.

For the PSG case the derivation of the gap potential can be found in
\citet{2013arXiv1304.4203S,2015MNRAS.447.2295S}, where they made the
same assumption as above, i.e., $B_\theta = B_{\phi} = 0$ and $B_r
\neq 0$, and for the PSG model, charge density can be written as $\rho
= (1 - \eta)\rho_{GJ} = (1-\eta)(\Omega \cdot B/2\pi c)$, where $\eta$
is the screening factor.  A spark develops in the PSG and has a height
$h_{\parallel}$ and width $h_{\perp}$.  So the angular size of the
spark can be approximated as $\Delta \theta = h_{\perp}/R$.  Under
these approximations and using Eq.~\ref{a2} and \ref{a4} in spherical
co-ordinates, \citet{2013arXiv1304.4203S} showed that $\Delta V$ can
be written in the form,

\begin{equation}
\frac{\Delta V}{h_{\parallel}^2} + \frac{\Delta V}{h_{\perp}^2} = \frac{2 \eta B_r \Omega \cos(\alpha +\mu)}{c} 
\label{a12}
\end{equation}

where the angle $\alpha + \mu$ is the angle between the local magnetic
field and the rotation axis.  RS75 assumed $h_{\perp} = h_{\parallel}
= h$ and $\eta =1 $, and an anti-aligned rotator $\alpha + \mu = 0$,
hence Eq.~\ref{a12} becomes the RS value $\Delta V = (B_r
\Omega/c)h^2$. For the PSG model however, the gap height
$h_{\parallel}$ needs to be much larger than the IVG case, since as
the potential in the gap is screened, higher potential drop is needed
for the full pair cascade to happen. However the spark width is
smaller than the height, i.e. $h_{\perp} \ll h_{\parallel}$. In this
condition the PSG gap potential $\Delta V_{psg}$ can be written as,

\begin{equation}
\Delta V_{psg} = \frac{4\pi \eta B_r \cos(\alpha + \mu)}{cP} h_{\perp}^2
\label{a13}
\end{equation}

Now one can write $E_{\theta} = \Delta V_{psg} / h_{\perp}$, and use
Eq.~\ref{adrift} to get an expression of the drift veloctiy
  $v_d^{\prime}$ in terms of $\eta$ as,

\begin{equation}
\vec{v}_d^{\prime} = - \frac{4 \pi \eta h_{\perp} \cos(\alpha +\mu)}{P} \hat{\phi}
\label{a12}
\end{equation}

the magnitude of which is the same as Eq. 3.50 of
\citet{2013arXiv1304.4203S}, and the negative sign indicates the
lagging behind nature of the sparks.

\bibliographystyle{mn2e}
\bibliography{J2144}
\end{document}